\documentclass[superscriptaddress,preprintnumbers,amsmath,amssymb,prd,nofootinbib,preprint]{revtex4-1}
\pdfoutput=1
\usepackage{graphicx}
\usepackage{epstopdf}
\usepackage{dcolumn}
\usepackage{bm}
\usepackage{hyperref}
\usepackage{color}
\usepackage{amsmath}
\usepackage{cancel}
\usepackage{xpatch}

\makeatletter
\def\l@subsubsection#1#2{}
\patchcmd{\@ssect@ltx}
    {\addcontentsline{toc}{#1}{\protect\numberline{}#8}}
    {}
    {}
    {}
\makeatother
\begin{document}

\def\a{\alpha}
\def\b{\beta}
\def\c{\varepsilon}
\def\d{\delta}
\def\e{\epsilon}
\def\f{\phi}
\def\g{\gamma}
\def\h{\theta}
\def\k{\kappa}
\def\l{\lambda}
\def\m{\mu}
\def\n{\nu}
\def\p{\psi}
\def\q{\partial}
\def\r{\rho}
\def\s{\sigma}
\def\t{\tau}
\def\u{\upsilon}
\def\v{\varphi}
\def\w{\omega}
\def\x{\xi}
\def\y{\eta}
\def\z{\zeta}
\def\D{\Delta}
\def\G{\Gamma}
\def\H{\Theta}
\def\L{\Lambda}
\def\F{\Phi}
\def\P{\Psi}
\def\S{\Sigma}

\def\o{\over}
\def\beq{\begin{align}}
\def\eeq{\end{align}}
\newcommand{\gsim}{ \mathop{}_{\textstyle \sim}^{\textstyle >} }
\newcommand{\lsim}{ \mathop{}_{\textstyle \sim}^{\textstyle <} }
\newcommand{\vev}[1]{ \left\langle {#1} \right\rangle }
\newcommand{\bra}[1]{ \langle {#1} | }
\newcommand{\ket}[1]{ | {#1} \rangle }
\newcommand{\EV}{ {\rm eV} }
\newcommand{\KEV}{ {\rm keV} }
\newcommand{\MEV}{ {\rm MeV} }
\newcommand{\GEV}{ {\rm GeV} }
\newcommand{\TEV}{ {\rm TeV} }
\newcommand{\1}{\mbox{1}\hspace{-0.25em}\mbox{l}}
\newcommand{\headline}[1]{\noindent{\bf #1}}
\def\diag{\mathop{\rm diag}\nolimits}
\def\Spin{\mathop{\rm Spin}}
\def\SO{\mathop{\rm SO}}
\def\O{\mathop{\rm O}}
\def\SU{\mathop{\rm SU}}
\def\U{\mathop{\rm U}}
\def\Sp{\mathop{\rm Sp}}
\def\SL{\mathop{\rm SL}}
\def\tr{\mathop{\rm tr}}
\def\mpl{M_{\rm Pl}}

\def\IJMP{Int.~J.~Mod.~Phys. }
\def\MPL{Mod.~Phys.~Lett. }
\def\NP{Nucl.~Phys. }
\def\PL{Phys.~Lett. }
\def\PR{Phys.~Rev. }
\def\PRL{Phys.~Rev.~Lett. }
\def\PTP{Prog.~Theor.~Phys. }
\def\ZP{Z.~Phys. }

\def\dd{\mathrm{d}}
\def\ff{\mathrm{f}}
\def\BH{{\rm BH}}
\def\inf{{\rm inf}}
\def\ev{{\rm evap}}
\def\eq{{\rm eq}}
\def\SM{{\rm sm}}
\def\Mpl{M_{\rm Pl}}
\def\GeV{{\rm GeV}}
\newcommand{\Red}[1]{\textcolor{red}{#1}}
\newcommand{\TL}[1]{\textcolor{blue}{\bf TL: #1}}


\title{
Minimal  Mirror Twin Higgs
}

\author{Riccardo Barbieri}
\affiliation{Institute of Theoretical Studies, ETH Zurich, CH-8092 Zurich, Switzerland}
\affiliation{Scuola Normale Superiore, Piazza dei Cavalieri 7, 56126 Pisa, Italy}
\author{Lawrence J. Hall}
\affiliation{Department of Physics, University of California, Berkeley, California 94720, USA}
\affiliation{Theoretical Physics Group, Lawrence Berkeley National Laboratory, Berkeley, California 94720, USA}
\author{Keisuke Harigaya}
\affiliation{Department of Physics, University of California, Berkeley, California 94720, USA}
\affiliation{Theoretical Physics Group, Lawrence Berkeley National Laboratory, Berkeley, California 94720, USA}

\begin{abstract}
 In a Mirror Twin World with a maximally symmetric Higgs sector the little hierarchy of the Standard Model can be significantly mitigated, perhaps displacing the cutoff scale above the LHC reach. We show that consistency with observations requires that the $Z_2$ parity exchanging the Standard Model with its mirror be broken in the Yukawa couplings.  A minimal such effective field theory, with this sole $Z_2$ breaking, can generate the $Z_2$ breaking in the Higgs sector necessary for the Twin Higgs mechanism.  The theory has constrained and correlated signals in Higgs decays, direct Dark Matter Detection and Dark Radiation, all within reach of foreseen experiments, over a region of parameter space where the fine-tuning for the electroweak scale is 10-50\%.  For dark matter, both mirror neutrons and a variety of self-interacting mirror atoms are considered.  Neutrino mass signals and the effects of a possible additional $Z_2$ breaking from the vacuum expectation values of $B-L$ breaking fields are also discussed. 
\end{abstract}

\date{\today}

\maketitle

\tableofcontents
\section{Introduction}

An intriguing idea, that has persisted over several decades, is that of the Mirror World: the Standard Model, with quarks and leptons $(q,l)$ and gauge interactions $SU(3) \times SU(2) \times U(1)$, is supplemented by an identical sector where mirror quarks and leptons $(q',l')$ interact via mirror gauge interactions $SU(3)' \times SU(2)' \times U(1)'$.  There are two prime motivations for this idea.  The discrete symmetry that interchanges the ordinary and mirror worlds can be interpreted as spacetime parity, $P$,  allowing a neat restoration of parity \cite{Lee:1956qn,Kobzarev:1966qya}.  Secondly, mirror baryons are expected to be produced in the early universe and to be sufficiently stable to yield dark matter, and this may lead to an understanding of why the cosmological energy densities of baryons and dark matter are comparable.

A third, more recent motivation for the Mirror World arises from the absence so far of new physics at colliders to explain the origin of the weak scale.  If the Higgs doublets of the two sectors $(H,H')$ possess a potential with maximal $SO(8)$ symmetry at leading order, the observed Higgs boson becomes a pseudo-Nambu-Goldstone boson with a mass that is insensitive to the usual Standard Model (SM) quadratic divergences; this is the Twin Higgs idea~\cite{Chacko:2005pe}.   Furthermore, if the symmetric quartic coupling of this potential, $\lambda$, is large relative to the SM quartic coupling, $\lambda_{\rm SM}$, the Mirror World reduces the amount of fine-tuning by a factor of $ 2 \lambda / \lambda_{\rm SM}$ to reach any particular UV cutoff of the effective theory.  Since we now know that $\lambda_{\rm SM} = 0.13$ is small, this improvement can be very significant, allowing a Little Hierarchy between the weak scale and the UV cutoff, which may be beyond the LHC reach.   

In this paper we formulate a minimal, experimentally viable, low energy effective theory for this idea, Minimal Mirror Twin Higgs, and study its signals.   This is a pressing issue: mirror baryon dark matter, the Twin Higgs mechanism and consistency with cosmological limits on the amount of dark radiation all require a breaking of parity, $P$ \cite{Barbieri:2005ri}.  How is this to be accomplished?   We do not attempt a UV completion, whether supersymmetric~\cite{Chang:2006ra,Craig:2013fga} or with composite Higgs~\cite{Batra:2008jy,Geller:2014kta,Barbieri:2015lqa,Low:2015nqa}, but note that both account for the approximate $SO(8)$ symmetry of the Higgs potential.  

The $SO(8)$ invariant quartic interaction contains an interaction $H^\dagger H H'^\dagger H'$ thermally coupling the two sectors at cosmological temperatures above a few GeV, so that the bound on dark radiation provides a very severe constraint on Mirror Twin Higgs.   The Twin Higgs mechanism requires a parity breaking contribution to the Higgs mass terms in the potential, $\Delta m_H^2$.  We find this term by itself to be insufficient to solve the dark radiation problem, nomatter what other interactions connect the two sectors, at least for fine-tunings above the percent level.  This then implies that the Yukawa couplings of the two sectors differ, $y' \neq y$.

Hence we introduce an effective theory, Minimal Mirror Twin Higgs, where all $P$ violation arises from a breaking of flavor symmetry, yielding different Yukawa couplings in the two sectors.    This single source of $P$ violation leads simultaneously to three key results
\begin{itemize}
\item  The $\Delta m_H^2$ term necessary for the Twin Higgs mechanism is generated via $q'$ loops. 
\item The mirror QCD phase transition temperature is raised above the decoupling temperature of the two sectors,   
 solving the problem of excessive dark radiation.
\item The mirror baryon mass is raised, allowing viable dark matter.
\end{itemize}

A striking signature would be the discovery at the LHC, or a future collider, of the mirror Higgs itself,
decaying to $WW$ or $ZZ$; see \cite{Barbieri:2005ri} and Fig.~10 of \cite{Buttazzo:2015bka}.  As the mirror Higgs mass depends on the $SO(8)$ invariant quartic, $\lambda$, it could be beyond the LHC range, and here we focus on other signals.  The size of $P$ breaking in the Yukawa couplings to obtain the above three results leads to a preferred range of the lightest $q'$ mass of $(2-20)$ GeV, leading us to compute signals for the following quantities 
\begin{itemize}
\item The signal strength, $\mu$, and the invisible width, $\Gamma(h \rightarrow \rm inv)$, of the Higgs boson.
\item  The amount of dark radiation, $\Delta N_{\rm eff}$. 
\item The direct detection rate for mirror baryon dark matter from Higgs exchange.
\item The effective sum of neutrino masses affecting large scale structure and the CMB.
\end{itemize}
These signals are tightly correlated as they all depend on the Higgs portal between the two sectors, the ratio of the weak scales, and on the masses of the light $q'$.  For dark matter, both mirror neutrons and a variety of self-interacting mirror atoms are considered.

After a brief review of the Twin Higgs mechanism in section~\ref{sec:TH}, we
demonstrate that the breaking of $P$ in the Yukawa couplings is necessary in section~\ref{sec:Z2BYuk}.
We define the Minimal Mirror Twin Higgs theory in section~\ref{sec:MinEFT}, and discuss the consequences of breaking $P$ in the Yukawa couplings.
We constrain the $q'$ Yukawa couplings from $\Gamma(h \rightarrow \rm inv)$ and $\Delta m_H^2$, and study how large the mirror QCD phase transition temperature $T'_c$ can be.
In section~\ref{sec:cosmology} we study the cosmological history of the two sectors when the only communication between them arises from the Higgs interaction $H^\dagger H H'^\dagger H'$ and find that the decoupling temperature can be lower than $T'_c$, allowing $\Delta N_{\rm eff}$ to lie inside the observational limit.  We predict the amount of dark radiation and the effective sum of neutrino masses.  In section~\ref{sec:mixing} we examine an alternative cosmology when communication between the sectors is dominated by kinetic mixing of the hypercharge gauge bosons. We study a variety of candidates for mirror dark matter in section~\ref{sec:DM}, and find that the $H^\dagger H H'^\dagger H'$ interaction, together with the enhanced $q'$ Yukawa couplings, will allow direct detection at planned experiments over a large part of the mass range. In section~\ref{sec:heavynu'} we briefly study $\Delta N_{\rm eff}$ and dark matter candidates when additional $P$ breaking arises from the absence of Majorana masses for right-handed mirror neutrinos.   In the Appendix we show that a PQ symmetry common to both sectors allows a solution to the strong CP problem, with the axion mass enhanced by the mirror sector by a factor of order $10^3$, leading to the possibility that $f_a$ is of order 10 TeV.

\section{Review of the Twin Higgs mechanism}
\label{sec:TH}

In this section, we review the Twin Higgs mechanism using a linear sigma model.
The Standard Model Higgs doublet $H$, together with a scalar field $H'$,
is embedded into
a representation of some approximate global symmetry.
The global symmetry is broken down to a subgroup containing $SU(2)_L\times U(1)_Y$ by the vacuum expectation value (VEV) of $H'$.
Four of the pseudo-Nambu-Goldstone bosons form a doublet of $SU(2)_L$ and are identified with the Standard Model Higgs doublet.
The lightness of the Higgs mass
in comparison with the scale of the Higgs potential can be understood in this way.

Let us take a closer look at the Higgs potential.
A global symmetry  preserving potential is given by
\begin{align}
V_{\rm sym} =& \lambda \left( |H|^2 + |H'|^2 \right)^2 
+ m_{H}^2 \left( |H|^2 + |H'|^2 \right).
\label{eq:Higgs potential}
\end{align}
We have neglected possible higher order terms which are expected in composite Twin Higgs models, as they are irrelevant for the following discussion. 
Since the global symmetry is explicitly broken by Yukawa couplings and the electroweak gauge interaction, we expect a breaking of the global symmetry in the Higgs potential, at least by quantum corrections.
The quantum correction to the mass term is the most dangerous, and to suppress it
we assume a $Z_2$ symmetry $H\leftrightarrow H'$ and call $H'$ the mirror Higgs.
We also introduce appropriate mirrors of other SM particles.
In the following, we use `` $'$ " to denote mirror objects.
The global symmetry of the Higgs potential is now $SO(8)$.
A $Z_2$ symmetric mass term is accidentally $SO(8)$ symmetric,
and an $SO(8)$ breaking potential is given by%
\footnote{%
\small Quantum corrections from top quarks gives $\Delta \lambda = 3y_t^4/64\pi^2 (\log (\Lambda^2/m_t^2)+A)$, with $\Lambda$ the cut off scale of the Higgs sector and $A$ a finite UV-dependent term. To obtain the observed Higgs mass requires $\log(\Lambda/m_t)+A/2 \approx 7$. Alternatively a tree level $\Delta \lambda$ may exist, as, e.g., from a supersymmetric D-term. See \cite{Barbieri:2015lqa} for the case of a composite Twin Higgs model.}
\begin{align}
V_{\bcancel{SO(8)}} = \Delta \lambda \, (|H|^4 + |H'|^4).
\label{eq:su4breaking}
\end{align}
As we will see later, we need small $Z_2$ breaking terms to obtain a correct electroweak symmetry breaking scale.
We assume $Z_2$ breaking in the mass term,
\begin{align}
V_{\bcancel{Z_2}} = \Delta m_{H}^2 \, |H|^2.
\label{eq:z2breaking}
\end{align}
The origin of the $Z_2$ breaking mass term is explained in the next section.

Let us derive the VEVs of the Higgs fields in the small $SO(8)$ breaking limit.
Assuming $\Delta m^2_{H} >0$, we expect $\vev{H'}^2 > \vev{H}^2$.
Setting $\vev{H}=0$ initially, the VEV of $\vev{H'}$ is given by
\begin{align}
v'^2 \equiv& \vev{H'}^2 = \frac{-m_H^2}{2\lambda}.
\end{align}
In the unitary gauge of $SU(2)_L\times U(1)_Y$, the pseudo-Nambu Goldstone boson, namely the Standard Model like Higgs $h$, is given by
\begin{align}
\begin{pmatrix}
  H \\ H'
\end{pmatrix}
\rightarrow
v' 
\begin{pmatrix}
{\rm sin} \frac{h}{\sqrt{2}v'} \\  {\rm cos} \frac{h}{\sqrt{2}v'}
\end{pmatrix}.
\label{eq:ngb Higgs}
\end{align}
Minimizing the potential of $h$ given by Eqs.~(\ref{eq:su4breaking}) and (\ref{eq:z2breaking}), we obtain
\begin{align}
v^2 \equiv \vev{H}^2 = v'^2{\rm sin}^2 \frac{\vev{h}}{\sqrt{2} v'} \simeq  \frac{v'^2}{2}
\left(
1 - \frac{\Delta m_H^2}{2 \Delta \lambda v'^2}
\right).
\label{eq:v}
\end{align}
The mass of $h$ is given by
\begin{align}
m_h^2 \simeq 8 \Delta \lambda \; v^2,
\label{eq:Higgs mass}
\end{align}
whereas the mass of the mirror Higgs boson, $h'$, is
\begin{align}
m_{h'}^2 \simeq 4  \lambda \; v'^2.
\label{eq:mirror Higgs mass}
\end{align}
From Eqs.~(\ref{eq:v}) and (\ref{eq:Higgs mass}), the required $Z_2$ breaking is given by
\begin{align}
\Delta m_H^2  \simeq  \frac{1}{4} \frac{v^{'2}}{v^2} m_h^2 \simeq (200~{\rm GeV})^2 \times \left( \frac{v'/v}{3} \right)^2.
\label{eq:Z2breaking_mass}
\end{align}

Let us comment on the fine-tuning in the Twin Higgs model.
In order to obtain the hierarchy between $v'$ and $v$, $\Delta m_H^2$ must be tuned against $\Delta \lambda \, v'^2$.
The standard fine-tuning measure is given by
\begin{align}
\Delta \equiv \Bigl| \frac{\partial \, {\rm ln}v^2}{\partial \, {\rm ln} \Delta m_H^2}  \Bigr| = \frac{1}{2} \frac{v'^2}{v^2}.
\label{eq:Delta mH}
\end{align}
The mixing in the physical higgs bosons, $h$ and $h'$, imply an overall reduction of the couplings of the Standard Model higgs to any Standard Model particles by the relative amount $(1-1/2 (v/v')^2)$. The precision measurements of these couplings~\cite{Khachatryan:2016vau} require that $v'/v > 2$ at $95 \%$ C.L. (see section~\ref{sec:higgs_decay})%
\footnote{%
\small The electroweak precision measurement as well sets an indirect bound on $v'/v$, which depends on the mass of the mirror Higgs. For $m_{h'} = 1$ TeV, considering the IR contribution only, one obtains $v'/v > 3$ at $90\%$ C.L.}.
Hence we need a tuning of at least $50\%$: this is the unavoidable, minimal fine-tuning in Twin Higgs models.

In general we expect a fine tuning also in the mass of the mirror Higgs boson. Assuming that the dominant contribution to this fine tuning comes from the top loop suitably cutoff at a scale $\Lambda_{\rm TH}$
\footnote{%
\small For $\lambda> 1$ this requires a suppression of the Higgs loop contribution in the UV completion of the Twin Higgs model considered here.}, it is
\begin{equation}
\Delta_{m_{h'}} \approx \frac{3/8\pi^2 y_t^2 \Lambda_{\rm TH}^2}{2\lambda v^{'2}}.
\label{eq:tune_TH}
\end{equation}
If $\Delta_{m_{h'}}> 1$, the overall fine tuning in Twin Higgs is given by 
\begin{equation}
\Delta_{\rm TH} = \frac{1}{2}\frac{v'^2}{v^2} \times \Delta_{m_{h'}},
\end{equation}
that can be compared with the fine tuning in the SM
\begin{equation}
\Delta_{\rm SM} = \frac{3/8\pi^2 y_t^2 \Lambda_{\rm SM}^2}{2\lambda_{\rm SM}v^2},
\label{eq:tune_SM}
\end{equation}
where $\lambda_{\rm SM}$ is the SM quartic coupling and $\Lambda_{\rm SM}$ is the cut off scale of the SM.

Thus the fine tuning in TH relative to the SM is
\begin{equation}
\Delta_{\rm SM}/ \Delta_{\rm TH} = \frac{2\lambda}{ \lambda_{\rm SM}}\frac{\Lambda_{\rm SM}^2}{\Lambda_{\rm TH}^2}.
\label{eq:relative}
\end{equation}
As $\lambda_{\rm SM}\simeq 0.13$, this improvement is significant: for a moderate tuning $\Lambda_{\rm TH}$ can be above the scales directly explorable at the LHC.

Let us comment on the required quality of the $Z_2$ symmetry~\cite{Craig:2015pha,Barbieri:2015lqa}.
The top Yukawa coupling $y_t$ gives a one-loop quantum correction,
\begin{align}
\Delta m_H^2|_{y_t} \simeq \frac{3}{8\pi^2} \left(y_{t'}^2 - y_t^2\right ) \Lambda^2,
\label{eq:top yukawa_z2}
\end{align}
where $\Lambda$ is the cut off of the Higgs sector.
In composite Twin Higgs models, $\Lambda$ is the scale of higher resonances, which is expected be as large as $\Lambda\sim g_* v'$, where $g_*$ is the coupling strength of hadrons.
The naive-dimensional analysis~\cite{Manohar:1983md,Luty:1997fk} and the large $N$ counting~\cite{'tHooft:1973jz} suggest $g_*\sim 4\pi/\sqrt{N}$, where $N$ is the size of the confining gauge group of the composite Twin Higgs model.
In the supersymmetric Twin Higgs model, $\Lambda$ is the stop mass scale multiplied by a log enhancement factor.
Requiring that this correction does not exceed the required one in Eq.~($\ref{eq:Z2breaking_mass}$), we obtain
\begin{align}
\Bigl|\frac{y_{t'} - y_t}{y_t}\Bigr| \lesssim 0.03 \left( \frac{v'/v}{3} \right)^2 \left( \frac{5~{\rm TeV}}{\Lambda} \right)^2.
\label{eq:top yukawa}
\end{align}
The strong interaction gives a two loop quantum correction,
\begin{align}
\Delta m_H^2|_{\rm strong} \simeq \frac{3 y_t^2 \left( g_3^2 - g_3'^2 \right) }{8\pi^4} \Lambda^2,
\end{align}
leading to the requirement
\begin{align}
\Bigl|\frac{g_3'^2 - g_3^2}{g_3^2}\Bigr| \lesssim  0.5 \left( \frac{v'/v}{3} \right)^2 \left( \frac{5~{\rm TeV}}{\Lambda} \right)^2.
\label{eq:strong}
\end{align}
Finally, the $SU(2)_L$ interaction gives a one-loop quantum correction,
\begin{align}
\Delta m_H^2|_{\rm weak} \simeq \frac{9 \left( g_2^2 - g_2'^2 \right) }{64\pi^2} \Lambda^2,
\end{align}
requiring
\begin{align}
\Bigl|\frac{g_2'^2 - g_2^2}{g_2^2}\Bigr| \lesssim  0.2 \left( \frac{v'/v}{3} \right)^2 \left( \frac{5~{\rm TeV}}{\Lambda} \right)^2.
\label{eq:weak}
\end{align}

A natural explanation for the quality of the $Z_2$ symmetry shown in Eqs.~(\ref{eq:top yukawa}), (\ref{eq:strong}) and (\ref{eq:weak}) is that the Lagrangian is precisely $Z_2$ symmetric, with a complete copy of all Standard Model particles  -- this is nothing but the Mirror World.  A key question then becomes the form and origin of the $Z_2$ breaking necessary to construct a fully realistic theory.

\section{Necessity of $Z_2$ symmetry breaking in Yukawa couplings}
\label{sec:Z2BYuk}

As reviewed in the previous section, the Twin Higgs mechanism requires a $Z_2$ symmetry under which the Standard Model and mirror particles are exchanged.
The $Z_2$ symmetry must be broken eventually to obtain the correct electroweak symmetry breaking scale.
In this section, we show that it is mandatory to break the $Z_2$ symmetry in Yukawa couplings to suppress the abundance of dark radiation,
no matter what  the origin of $Z_2$ breaking is and no matter what interactions might be added to the theory to couple the two sectors.
Thus the minimal phenomenologically viable way to break the $Z_2$ symmetry is via $y_{f'} \neq y_f$.

The Standard Model and mirror particles interact with each other by the interaction in Eq.~(\ref{eq:Higgs potential}) so that
in the early universe at temperatures of order the weak scale the two sectors are kept in equilibrium via Higgs exchange.  At lower temperatures the mirror particles eventually annihilate/decay into mirror photons (and mirror neutrinos), giving
an extra component of relativistic particles, which is often referred to as dark radiation.  The amount of dark radiation depends on the decoupling temperature between the two sectors, $T_d$, below which the interaction rate between Standard Model and mirror particles is smaller than the Hubble expansion rate. Without introducing extra interactions, $T_d$ is determined by Higgs exchange which depends on the Yukawa couplings~\cite{Barbieri:2005ri, Chang:2006ra}.
In this section, to be general, we allow additional interactions coupling the sectors and treat $T_d$ as a free parameter.

\begin{figure}[t]
\centering
\includegraphics[clip,width=.48\textwidth]{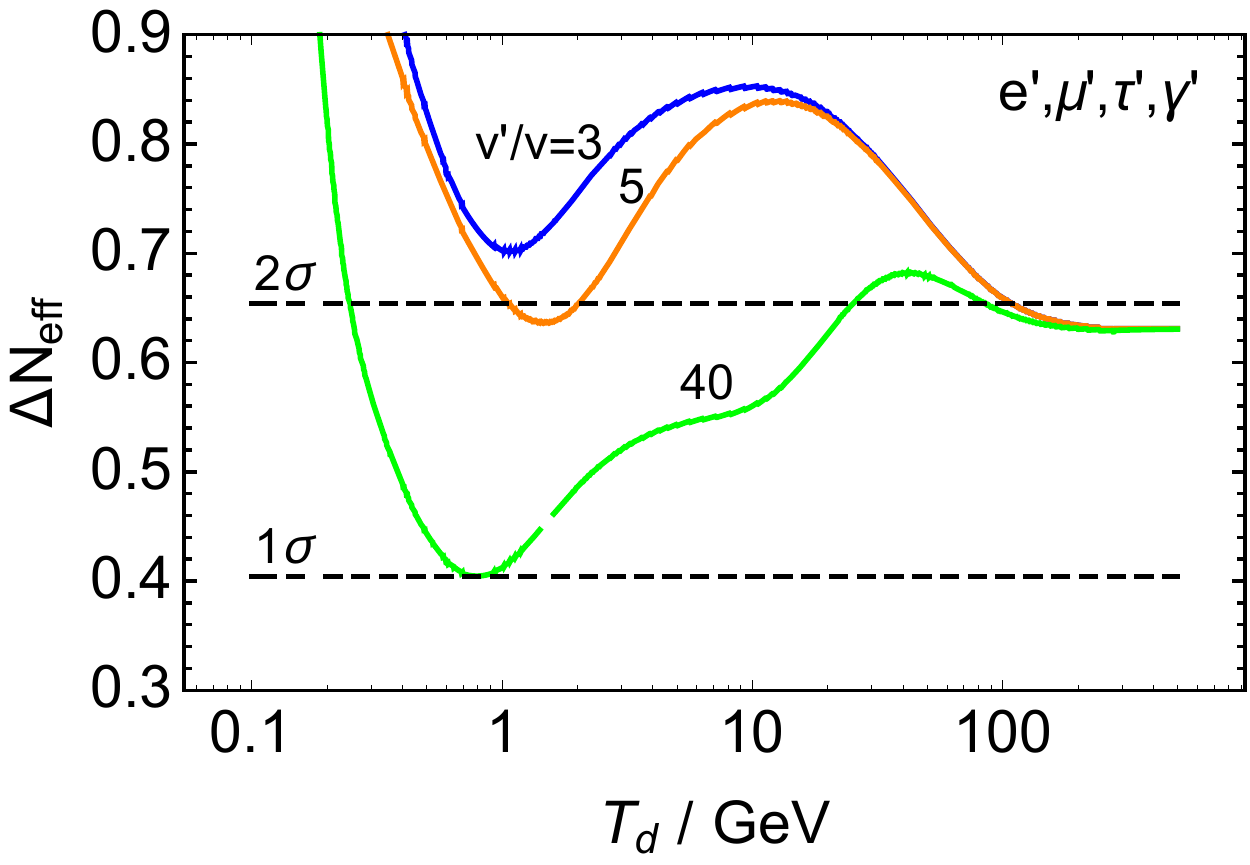}
\caption{Prediction for the amount of extra dark radiation arising from only $e'$, $\mu'$, $\tau'$ and $\gamma'$ as a function of the decoupling temperature $T_d$ between the Standard Model and mirror sectors.
}
\label{fig:Neff_lepton}
\end{figure}

Assuming the Yukawa couplings are $Z_2$ symmetric, the mirror charged lepton masses are determined solely by the ratio $v'/v$.
(The masses of mirror hadrons could be also affected by the $Z_2$ symmetry breaking in the $SU(3)_c$ gauge coupling.)
As their masses are relatively small, they remain in the thermal bath until low temperature, and contribute too much dark radiation.
In Fig.~\ref{fig:Neff_lepton}, we show the prediction for the abundance of dark radiation, which by convention is expressed as an effective extra number of neutrinos, $\Delta N_{\rm eff}$, which is given by
\begin{align}
\label{eq:DeltaNeff}
\Delta N_{\rm eff} = \frac{4}{7}g'_r \times \left( \frac{10.75}{g(T_d)} \right)^{4/3} \times \left(  \frac{g'(T_d)}{g'_r} \right)^{4/3},
\end{align}
where $g(T)$ and $g'(T)$ are the effective entropy degrees of freedom (d.o.f) of the Standard Model particles and the mirror particles at temperature $T$, respectively, and $g'_r$ is the d.o.f.~of the radiation component of the mirror sector.
The second factor expresses the heating of the Standard Model neutrinos, and the third factor expresses the heating of the dark radiation.
The d.o.f of the Standard Model particles $g(T)$ is extracted from~\cite{Borsanyi:2016ksw}.
We assume that only the mirror electron, muon, tau and photon are light and contribute to the dark radiation.
Light mirror neutrinos, as considered in sections \ref{sec:MinEFT} to \ref{sec:DM}, make the claim in this section even stronger.
Then $g'_r=2$ and
\begin{align}
g'(T_d) =& 2 + \frac{45}{2\pi^2} \frac{\rho'(T_d) + p'(T_d)}{T_d^4},\nonumber\\
\rho'(T_d) =& \frac{2}{\pi^2} \sum_{f=e,\mu,\tau}m_f^{4}\left(\frac{v'}{v}\right)^4 \int_1^\infty {\rm d}x\frac{\left(x^2-1\right)^{1/2}x^2}{{\rm exp}(xm_f v' /T_dv)+1},\nonumber\\
p'(T_d) =& \frac{2}{3\pi^2} \sum_{f=e,\mu,\tau}m_f^{4}\left(\frac{v'}{v}\right)^4 \int_1^\infty {\rm d}x\frac{\left(x^2-1\right)^{3/2}}{{\rm exp}(xm_f v' /T_dv)+1}.
\end{align}

The Planck collaboration puts a bound on the effective number of neutrinos, $N_{\rm eff}=3.2 \pm0.5$~\cite{Ade:2015xua}, which leads to the upper bound of $\Delta N_{\rm eff}<0.65~(2\sigma)$ and $0.40~(1\sigma)$,
indicated by dashed lines in Fig.~\ref{fig:Neff_lepton}.
The $2\sigma$ bound can be only marginally satisfied.
The $1\sigma$ bound can be satisfied when $v'/v \gtrsim40$, which requires a fine-tuning of more than $0.1$\% to obtain the electroweak symmetry breaking scale.
Note the importance to reach this conclusion of the deviation of the actual $g'(T_d)$ from a naive stepwise function.
We conclude that
it is necessary to break the $Z_2$ symmetry in the Yukawa couplings to further raise the mirror charged lepton masses.

The Yukawa couplings can naturally acquire $Z_2$ symmetry breaking if they arise from VEVs of fields, as in the Froggatt-Nielsen (FN) mechanism~\cite{Froggatt:1978nt}.
Once these fields take asymmetric VEVs, the $Z_2$ symmetry of the Yukawa couplings is spontaneously broken.
As long as the top Yukawa coupling does not depend on these field values, the Twin Higgs mechanism is maintained.
In this paper we do not specify the model of the FN mechanism, but study the physical consequences of a low energy effective field theory for Twin Higgs with $y_{f'} \neq y_f$.

\section{Minimal Mirror Twin Higgs}
\label{sec:MinEFT}

The effective field theory below $\Lambda_{\rm TH}$ for Minimal Mirror Twin Higgs is
\begin{align}
\mathcal{L}_{\rm EFT} &= \mathcal{L}_{321}(g_i, y_f, \lambda, m_H) + \mathcal{L}'_{3'2'1'}(g_{i'}, y_{f'}, \lambda', m_{H'}) + 2\lambda'' H^\dagger H H'^\dagger H' + \frac{1}{2} \frac{\epsilon}{{\rm cos}\theta_W^2} B^{\mu \nu} B'_{\mu \nu} \nonumber \\
&+ (LH)^2/M_1 + (L'H')^2/M_1 + (LH)(L'H')/M_2 + ....
\label{eq:EFT}
\end{align}
where $\mathcal{L}_{321}(g_i, y_f, \lambda, m_H)$ describes the Standard Model including all operators up to dimension 4 and $ \mathcal{L}'_{3'2'1'}$ similarly describes the mirror sector.  The lepton-Higgs interactions of the second line are the most general $Z_2$ symmetric set of dimension 5 operators and are suppressed by large masses $M_{1,2}$.  Including all operators consistent with gauge invariance allows kinetic mixing via $\epsilon$.   

The Twin Higgs mechanism is imposed by boundary conditions at $\Lambda_{\rm TH}$:
\begin{align}
\lambda'' = \lambda' = \lambda, \hspace{0.7in} m_{H'} = m_H, \hspace{0.7in} g_{i'} = g_i \hspace{0.7in}  y_{t'} = y_t.
\label{eq:bc1}
\end{align}
As discussed in footnote~1, there could also be a boundary condition giving non-zero $\Delta\lambda$.
$Z_2$ breaking is the minimal consistent with the requirement of the previous section, namely 
\begin{align}
y_{f'} \neq y_f \hspace{0.7in}  f \neq t.
\label{eq:bc2}
\end{align}  
This breaking of $Z_2$ is hard, meaning that the boundary conditions of (\ref{eq:bc1}) are not exact and are broken by loop corrections at $\Lambda_{\rm TH}$.
We restrict the size of $y_{f'}$, for $f \neq t$, so that these corrections maintain the Twin Higgs mechanism and satisfy Eqs.~(\ref{eq:top yukawa}, \ref{eq:strong}, \ref{eq:weak}).
The dimension 5 operators have flavor structure suppressed and would arise from the seesaw mechanism with $Z_2$ symmetric neutrino Yukawa couplings and right-handed neutrino masses.  Comparable $M_{1,2}$ are excluded from limits on oscillation to sterile neutrinos, suggesting that lepton number symmetries (or $B-L$ symmetries) play an important role in neutrino masses\footnote{\small If the hierarchy between $M_1$ and $M_2$ is not too large, the theory may have neutrino oscillation signals.}. 
We assume these symmetries lead to neutrino couplings $LNH + L'N'H'$ and masses for right-handed neutrinos, $N$ and $N'$, so that $M_2 \gg M_1$ or $M_1\gg M_2$; the light neutrinos are then Majorana or Dirac, respectively.

For $M_2 \gg M_1$, arising if $NN'$ mass mixing is absent, neutrinos are Majorana with
light mirror neutrino masses given by
\begin{align}
m_{\nu'} = (v'/v)^2 m_\nu.
\label{eq:mnuM}
\end{align}
For $M_1 \gg M_2$, arising if the only right-handed neutrino mass is $NN'$,
Standard Model and mirror left-handed neutrinos obtain Dirac masses, so that
\begin{align}
m_{\nu'} =  m_\nu.
\label{eq:mnuD}
\end{align}
Such Dirac masses from a seesaw are an interesting possible consequence of parity restoration in the mirror scheme.

Next we constrain the $Z_2$ breaking in $y_{f'}$ by requiring that quantum corrections to the Higgs masses yield non-zero $\Delta m_H^2$ as required for the Twin Higgs mechanism in Eq.~(\ref{eq:Z2breaking_mass}).  We then place experimental bounds on $y_{f'}$ from the invisible decay width of the SM-like Higgs.   Finally, we study mirror fermion spectra, consistent with these constraints, that maximize the QCD$'$ scale.

\subsection{Constraint on Yukawa couplings from $\Delta m_H^2$}

We derive a constraint on $Z_2$ symmetry breaking in the Yukawa couplings
by requiring that quantum correction to the soft Higgs masses yields the Twin Higgs mechanism.
We assume the top Yukawa couplings of the Standard Model and mirror sectors are identical
but allow other Yukawa couplings to differ, inducing a $Z_2$ breaking Higgs mass term
\begin{align}
\Delta m_H^2|_{\rm Yuk} \, \simeq \, \sum_{f \neq t} \frac{N_{f'}}{8\pi^2} \, y_{f'}^2(\Lambda) \, \Lambda^2,
\label{eq:}
\end{align}
where $N_{f'}$ is the multiplicity of the mirror fermion $f'$; 3 for mirror quarks and one for mirror leptons.
It should be remarked that the sign of the mass term is the required one.
For this correction to explain $\Delta m_H^2$ of Eq.~(\ref{eq:Z2breaking_mass}),
the sum of the square of the mirror Yukawa couplings, and hence the mirror fermions masses, are determined
\begin{align}
 \label{eq:bound_yukawa}
 \sum_{f\neq t} \left( \frac{N_{f'}}{3} \right) y_{f'}^{2}(\Lambda) \; & \simeq  \; 0.04  \left( \frac{v'/v}{3} \right)^2 \left( \frac{5~{\rm TeV}}{\Lambda} \right)^2, \nonumber \\
  \sum_{f=u,d,c,s,b} \left( \frac{N_{f'}}{3 \, \delta_{f',\Lambda}^2 } \right) \, m_{f'}^{2} \; & \simeq  \; (100~{\rm GeV})^2 \left( \frac{v'/v}{3} \right)^4 \left( \frac{5~{\rm TeV}}{\Lambda} \right)^2.
\end{align}
Here, $\delta_{f',\mu} \equiv y_{f'}(m_{f'}) / y_{f'}(\mu)$ encodes the effect of the renormalization between a scale $\mu$ and $m_{f'}$.
$\delta_{q',\Lambda}$ is about $1.3-1.5$ for $m_{q'} =(100-10)$ GeV, and $\delta_{l',\Lambda}\simeq 1$.
In the following we take $\delta_{q',\Lambda}=1.4$.
Note that the estimation of $\Delta m_H^2|_{\rm Yuk}$ is UV sensitive and hence involves $O(1)$ uncertainty,
which we formally treat by varying the value of $\Lambda$ in Eq.~(\ref{eq:bound_yukawa}).
The $Z_2$ breaking correction to the Higgs quartic couplings is proportional to $y_{f'}^{4}$ and is negligible.

\subsection{ Standard Model like Higgs decays}
\label{sec:higgs_decay}
The Standard Model like Higgs $h$ is an admixture of the two doublets $H$ and $H'$, $h=c_\gamma H + s_\gamma H'$, with $s_\gamma^2 \equiv \sin^2\gamma = (v/v')^2$ up to corrections of order $(v/v')^4$ and $(m_h/m_{h'})^2$. 
In turn this leads to a universal reduction by a factor $c_\gamma$ of the Higgs couplings to all pairs of SM particles as well as to a coupling of the same Higgs  to 
 the mirror fermions
\begin{align}
{\cal L} \supset - y_{f'} \; H' f_L' \bar{f}_R'  \; \rightarrow \; - \frac{v}{\sqrt{2}v'} \, y_{f'} \; h f_L' \bar{f}_R' \; = \; -   \frac{v m_{f'}}{\sqrt{2}v^{'2} \delta_{f',m_h}} \; h f_L' \bar{f}_R',
\end{align}
where the QCD running factor for a mirror quark $f'=q'$, $\delta_{q',m_h}$, is about $1.3$ for $m_{q'}$ around 10 GeV,
and the precise value depends on the details of the mirror quark spectrum\footnote{\small In the following, we take $\delta_{q',m_h}=1.3$ for simplicity. This simplification essentially does not change the following discussion.}.

\begin{figure}[t]
\centering
\includegraphics[clip,width=.48\textwidth]{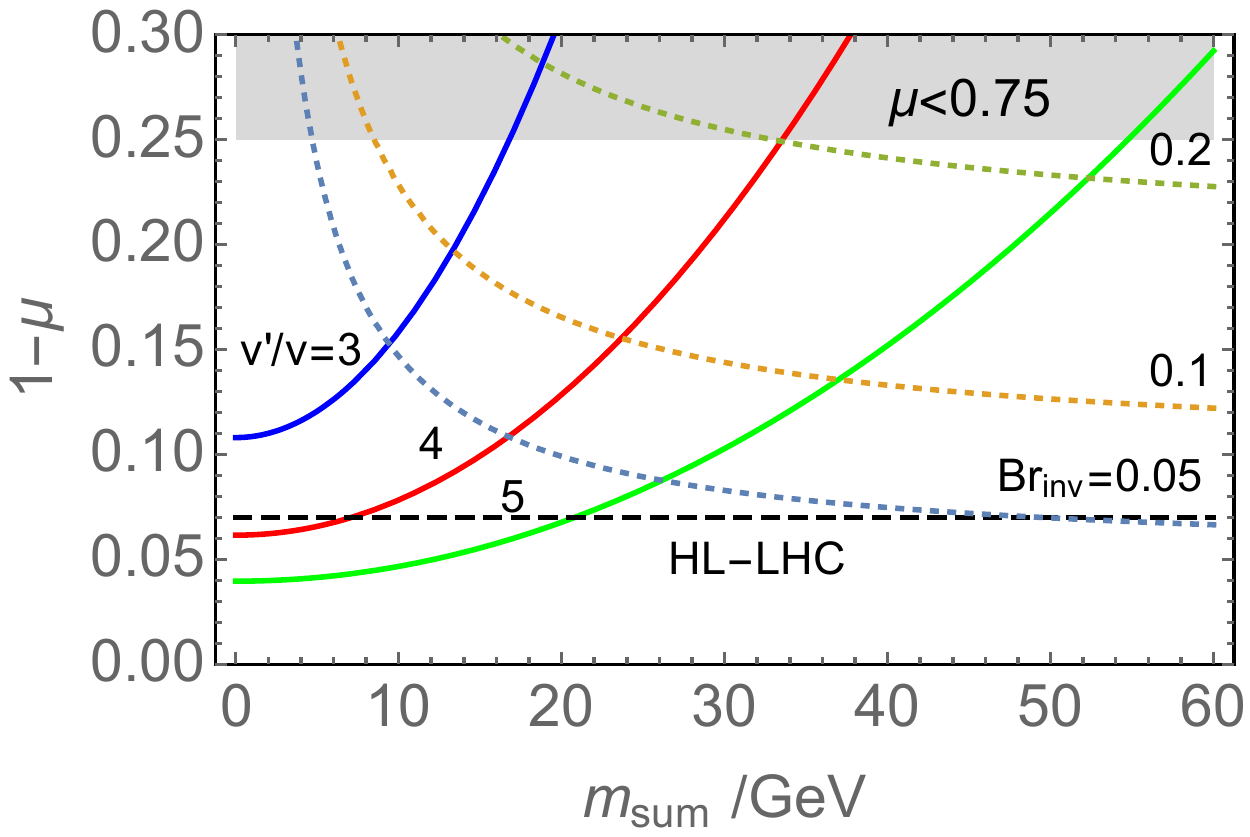}
\caption{
Prediction for the universal Higgs signal-strength as a function of a sum of mirror fermion masses defined in (\ref{eq:msum}), for given $v'/v$.
The dotted lines show contours of the invisible branching ratio, and the dashed line shows the sensitivity of the high-luminosity running of the LHC.
}
\label{fig:mass_mu}
\end{figure}

Higgs decays to mirror fermions leads to an invisible branching ratio
\begin{equation}
{\rm Br}_{\rm inv} = {\rm Br}(h\rightarrow f' \bar{f}') \simeq  0.1 \times \left( \frac{3}{v'/v} \right)^4 \sum_{f',2m_{f'}<m_h}  \frac{N_{f'}}{3}(\frac{m_{f'}}{10 {\rm GeV}})^2 
\delta_{f',m_h}^{-2}.
\end{equation}
where phase space has been neglected.  Fig.~\ref{fig:mass_mu} shows the correlations between $v/v'$,  ${\rm Br}_{\rm inv}$, and the universal deviation from unity of the Higgs signal-strengths at the LHC into any SM final state,
\begin{equation}
1-\mu = 1- c_\gamma^2(1-{\rm Br}_{\rm inv})\simeq s_\gamma^2 +{\rm Br}_{\rm inv},
\end{equation}
versus the relevant combination of mirror fermion masses
\begin{equation}
\sum_{f',2m_{f'}<m_h} m_{f'}^2 \frac{N_{f'}}{3}\left(\frac{1.3}{\delta_{f',m_h}}\right)^2  \equiv m^2_{\rm sum}.
\label{eq:msum}
\end{equation}
We adopt the bound $\mu>0.75$ (95\% C.L.) for the gluon fusion channel~\cite{Khachatryan:2016vau}, as it has the smallest uncertainty.
The bound is so strong that mirror fermions with $m_{f'} < m_h/2$
give sub-dominant contributions to $\Delta m_H^2$ (see Eq.~(\ref{eq:bound_yukawa})):
there must be at least one mirror fermion heavier than $m_h/2$ other than $t'$.
The high-luminosity LHC can probe $\mu<0.93$~\cite{ATLAS_prospect}, which is shown by a dashed line in Fig.~\ref{fig:mass_mu}.
Irrespective of the mirror fermions masses, $v'/v<4$ can be probed.
The ILC is expected to measure the Higgs signal-strength with an accuracy of $1\%$~\cite{Asner:2013psa}, which probes $v'/v<10$.

\subsection{Mirror QCD phase transition temperature}
\label{sec:Tc}
The larger Yukawa couplings of the mirror quarks leads to a larger mass of the mirror quarks, and hence to a larger mirror QCD phase transition temperature, $T_c'$.  Since the number of degrees of freedom of the mirror sector changes rapidly near the phase transition, increasing $T_c'$ above the decoupling temperature of the two sectors is critical to obtaining $\Delta N_{\rm eff}$ below the current bound, as specifically illustrated in section~\ref{sec:DR}.  However, there is a limit to how much $T_c'$ can be increased, and here we derive an upper bound on $T_c'$.

\begin{figure}[t]
\centering
\includegraphics[clip,width=.48\textwidth]{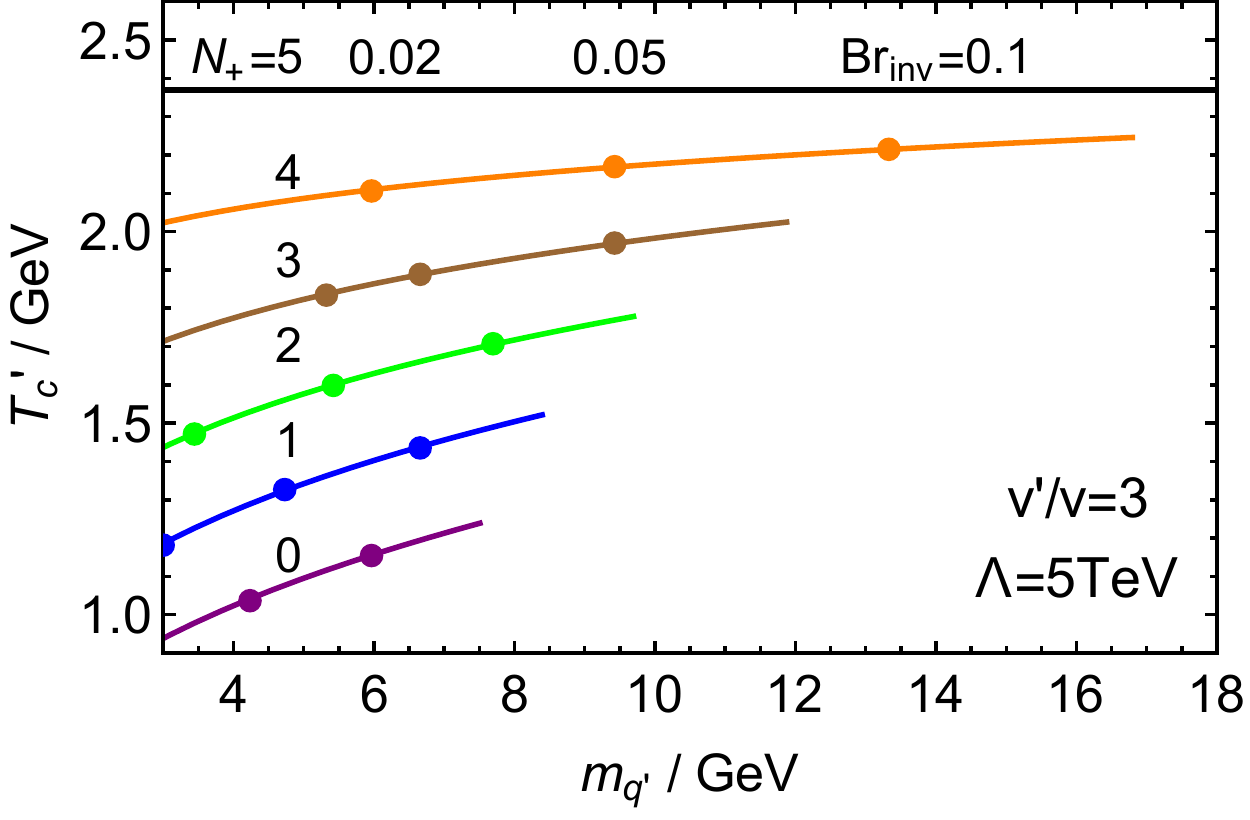}
\includegraphics[clip,width=.48\textwidth]{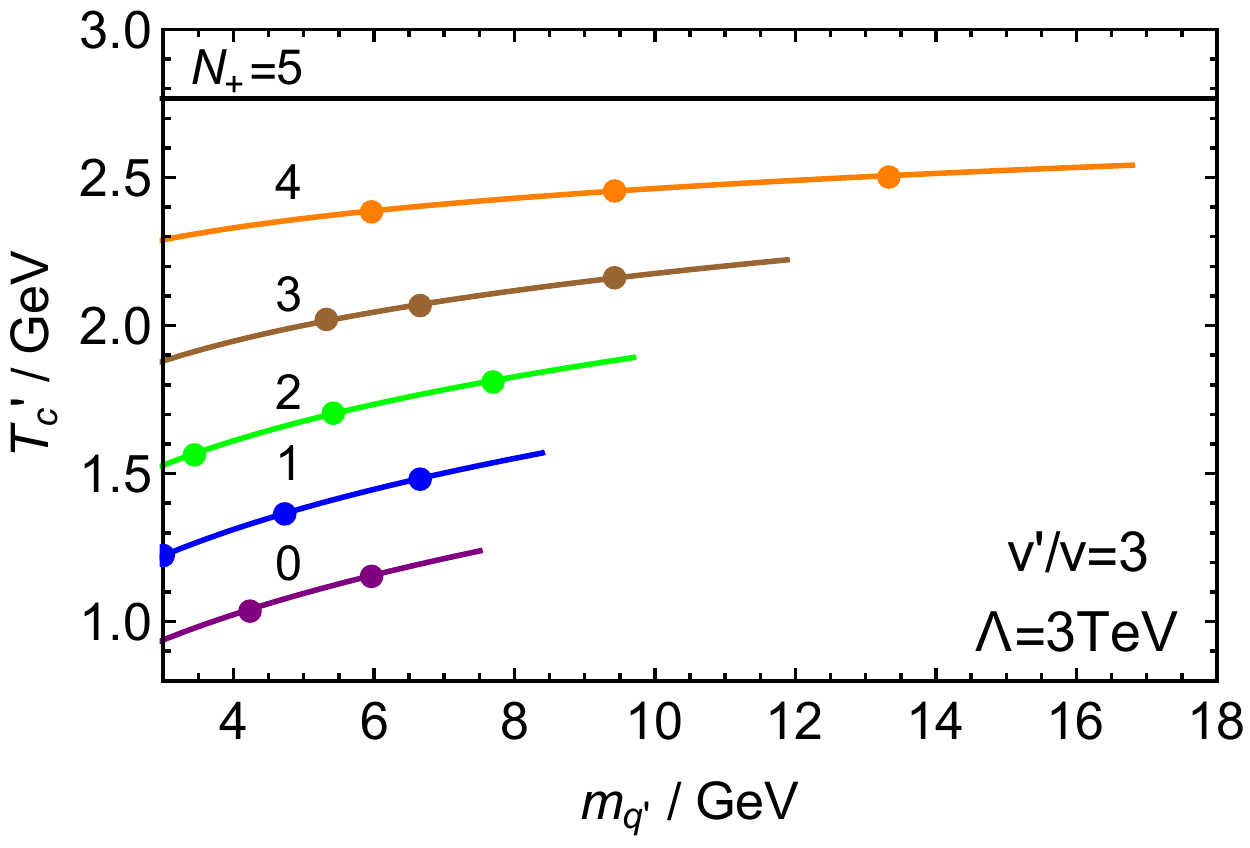}
\caption{
The prediction of the mirror QCD phase transition temperature as a function of the light mirror quark mass.
}
\label{fig:Tc'}
\end{figure}

First, we take the masses of $N_+$ mirror quarks ($m_+$) above $m_h/2$ and degenerate, and determine their mass according to Eq.~(\ref{eq:bound_yukawa}). Next, we take the masses of the remaining $5-N_+$ mirror quarks to be the same ($m_{q'}$), and constrained by the bound $\mu>0.75$.
With these masses, we solve the two-loop renormalization group equation of the mirror QCD gauge coupling constant.
We find that, for a wide range of parameter space, $m_{q'}$ is well above the scale at which the gauge coupling diverges.
Adopting the lattice calculation for the pure $SU(3)$ Yang-Mills gauge theory in Ref.~\cite{Okamoto:1999hi}, matching the renormalization scale and the inverse of the lattice spacing, we obtain $T_c'$, as shown in Fig.~\ref{fig:Tc'}.
The figure shows that $T_c'$ can be well above 1 GeV, and reach a few GeV.
Here the right end of each line shows the point where the bound from the invisible decay of the Higgs, as inferred from the limit on $\mu$, is saturated.
We also show values of ${\rm Br}(h\rightarrow{\rm inv})=0.02, 0.05, 0.1$ by dots on each line, from left to right.

Note that the sum of the square of Yukawa couplings is bounded as Eq.~(\ref{eq:bound_yukawa}) and $\mu > 0.75$,
while the dynamical scale of mirror QCD depends on the product of the mirror quark masses.
Thus, for a given $N_+$, the universal $m_+$ and $m_-$ saturating these bounds maximize $T_c'$:
the maximal $T_c'$ can be read off from the right hand end 
of each line.

\section{Thermal history with Higgs exchange}
\label{sec:cosmology}

In this section, we discuss the thermal history of the Minimal Mirror Twin Higgs theory of (\ref{eq:EFT}) in the limit of $\epsilon < 10^{-5}$, so that the  sectors are coupled only via Higgs exchange, and focus on the amount of dark radiation.  
Note that radiative corrections to $\epsilon$ in the effective theory below $\Lambda_{\rm{TH}}$  vanish at 3 loops%
\footnote{%
\small Candidates for such three-loop corrections are diagrams where a SM photon is connected with two Higgs with a loop of SM particles, and a mirror photon is connected with two mirror Higgs in a similar manner, and the two Higgs connect to two mirror Higgs via Higgs mixing. We find that the sub-diagram involving a photon and two Higgs vanishes due to the Bose statistic of the Higgs.  Intuitively speaking, the symmetrized two Higgs have vanishing angular momentum, and vanishing correlators with a photon.
},
and any 4-loop contribution would be of order $10^{-10}$.

In the early universe, at temperatures larger than several GeV, the Standard Model and mirror particles interact with each other and have the same temperature.
Below some temperature, $T_d$, the interaction between the sectors becomes inefficient and they evolve independently.
Heavy mirror particles eventually decay or annihilate into mirror photons and neutrinos, which are observed as dark radiation estimated in Eq.~(\ref{eq:DeltaNeff}).
To satisfy observational constraints requires $g'(T_d) \ll g(T_d)$ so that at $T_d$ the colored states $u,d,s,g$ contribute to $g(T_d)$ but QCD$'$ states contribute very little to $g'(T_d)$; roughly speaking, decoupling must occur between QCD$'$ and QCD phase transitions, and we explore this further below.

\subsection{QCD$'$ and constraints on the $q'$ spectrum}

If some $q'$ are light compared to the QCD$'$ scale $\Lambda'$, then $T_c'$ is lower than computed in the previous section resulting in large QCD$'$ contributions to $g'(T_d)$ that are excluded by bounds on $\Delta N_{\rm eff}$.  
Hence the QCD$'$ phase transition is purely gluonic with zero $q'$ flavors.   The temperature dependence of this zero flavor QCD, $g'_{QCD}(T)$ has been accurately computed on the lattice~\cite{Borsanyi:2012ve}.  
As the temperature decreases from high values, $g'_{QCD}(T)$ drops from its large perturbative value of 16 only very gradually, and then sharply drops very close to the critical temperature $T_c'$; immediately after the phase transition at $T_c'$ the glueball contribution to $g'(T_c')$ is 0.6.  If $T_d > 1.1 \, T_c'$, much of the entropy of the mirror gluon plasma is distributed solely to the mirror particles, which leads to too large $\Delta N_{\rm eff}$.
Hence, in the next sub-section we seek regions of parameter space where $T_d < T_c'$, which is at most (1--2.8) GeV, as shown in Fig.~\ref{fig:Tc'}.

Below $T_c'$, mirror glueballs $S'$ decay to $\gamma' \gamma'$ via
\begin{align}
{\cal L} \sim \frac{\alpha_s \alpha}{m_{q'}^4} \, G'G' F'F',
\end{align}
generated by integrating out the lightest mirror quark, of mass $m_{q'}$, giving a decay rate
\begin{align}
\Gamma({\cal  S}'\rightarrow \gamma' \gamma') = \frac{\alpha_s^2 \alpha^2}{64\pi^3} \frac{\Lambda^{'9}}{m_{q'}^8} \sim 10^{-17}~{\rm GeV} \left( \frac{\Lambda'}{2~{\rm GeV}} \right)^9 \left( \frac{20~{\rm GeV}}{m_{q'}} \right)^8.
\label{eq:glueball_decay}
\end{align}
$\Lambda'$ is the scale at which QCD$'$ becomes strongly coupled, and is comparable to $T_c'$.
In the case with all $q'$ heavier than $m_h/2$, labelled $N_+=5$ in Fig.~\ref{fig:Tc'}, the mirror glueball is not in thermal equilibrium just above $T_c'$ where its contribution to $g'$ is large.  In this case,  the mirror glueballs decay late, well after $T_d$ computed below, and hence is excluded by the limit on $\Delta N_{\rm eff}$.  
Fig.~\ref{fig:Tc'} shows that, in all the remaining cases of $N_+=1\mathchar`-4$, there must be one $q'$ lighter than 22 GeV.  Over much of the parameter space of the $q'$ spectrum, the decay rate of (\ref{eq:glueball_decay}) is fast enough to ensure that the glueballs at $T_c'$ are indeed in thermal equilibrium with $\gamma'$, and contribute 0.6 to $g'(T_c')$.
Cases with $\Gamma({\cal  S}'\rightarrow \gamma' \gamma')$ less than the Hubble expansion rate at $T_c'$ must be discarded as they give too much dark radiation.

Mirror glueballs also decay to $\nu'\bar{\nu}'$ via
\begin{align}
{\cal L} \sim \frac{\alpha_s \alpha_2}{m_{q'}^2 m_{Z'}^2} \, \nu^\dag \sigma \nu {\rm D}G'G',
\end{align}
giving a decay rate
\begin{align}
\Gamma({\cal  S}'\rightarrow \nu' \bar{\nu}') = \frac{\alpha_s^2 \alpha_2^2}{64\pi^3} \frac{\Lambda^{'9}}{m_{q'}^4 m_{Z'}^4} \sim 10^{-20}~{\rm GeV} \left( \frac{\Lambda'}{2~{\rm GeV}} \right)^9 \left( \frac{20~{\rm GeV}}{m_{q'}} \right)^4 \left( \frac{3}{v'/v} \right)^4,
\label{eq:glueball_decay_nu}
\end{align}
which is negligible in comparison with  $\Gamma({\cal  S}'\rightarrow \gamma' \gamma')$.

\subsection{Decoupling temperature}

Now, let us estimate the decoupling temperature $T_d$.
Mirror fermions $f'$ and standard fermions $f$ interact with each other by the exchange of the standard model Higgs.
Since  $f'$ interact with $\gamma'$, thermal equilibrium between $\gamma'$ and Standard Model particles can be maintained, even if $f'$ is heavy and its number density is much smaller than that of relativistic particles.
We discuss the thermal equilibrium of mirror neutrinos later.

Let us first estimate the interaction rate assuming that the dynamics of $f'$ in the thermal bath is described by that of free fermions.
This is certainly correct for the mirror leptons.
We comment on the case with mirror quarks later.
The scattering cross section between $f$ and $f'$ by Higgs exchange is given by
\begin{align}
\sigma v(ff'\rightarrow ff') = \frac{1}{8\pi}\left(\frac{m_f}{v}\right)^2 \left( \frac{v m_{f'}}{v^{'2}} \right)^2 \frac{m_f m_{f'}}{m_f + m_{f'}} \frac{p_{\rm cm}}{m_h^4}
\end{align}
where we assume a non-relativistic limit.
Here $p_{\rm cm}$ is the momentum of the fermion in the center of mass frame. In the thermal bath, it has a typical size
\begin{align}
p_{\rm cm}^2 = \frac{4T (m_f + m_{f'}+ \sqrt{m_f m_{f'}})}{3 \left( 2 + m_f/m_{f'} + m_{f'}/m_f\right)}.
\end{align}
The annihilation cross section of a pair of $f'$ into a pair of $f$ is given by
\begin{align}
\sigma(f'\bar{f}' \rightarrow f \bar{f})v = \frac{N_f}{4\pi} \left(\frac{m_f}{v}\right)^2 \left( \frac{v m_{f'}}{v^{'2}} \right)^2 \frac{ (m_{f'}^2 - m_f^2)^{3/2}}{m_{f'}^3m_h^4} p_{f'}^2.
\end{align}
Here $p_{f'}$ is the momentum of $f'$ in the center of mass frame. In the thermal bath, it is as large as $p_{f'}^2 \simeq 3m_{f'} T/2 $.
$N_f$ is the multiplicity of the Dirac fermion $f$; for one lepton (quark), $N_f = 1(3)$.

The energy density of mirror particles, $\rho'$, is transferred into Standard Model particles at a rate
\begin{align}
\frac{\rm d}{{\rm d}t}\rho' =& \sum_f 4 \left( N_f n_{F}\left(m_f,T\right)\right)  \left( N_{f'} n_{F}\left(m_{f'},T\right)\right)   \sigma v(ff'\rightarrow ff') \times \Delta E \nonumber \\
&+ \sum_f N_{f'}n_F(m_{f'},T)^2 \sigma v(f'\bar{f}'\rightarrow f\bar{f}) \times 2 m_{f'}.
\end{align}
Here $n_F(m,T)$ is the number density of a Weyl fermion of mass $m$ in the thermal bath at temperature $T$,
and $\Delta E \simeq T$ is a typical energy transfer by the scattering $f f' \rightarrow f f'$.
We find that the annihilation $f'\bar{f'}\rightarrow f \bar{f}$ is more important than the scattering $f f' \rightarrow f f'$ in determining $T_{d}$ by Higgs exchange.
This is because the energy transfer per annihilation is $2m_f'$ and is larger than that per scattering, $\Delta E \sim T$, for $T< m_{f'}$.

\begin{figure}[t]
\centering
\includegraphics[clip,width=.53\textwidth]{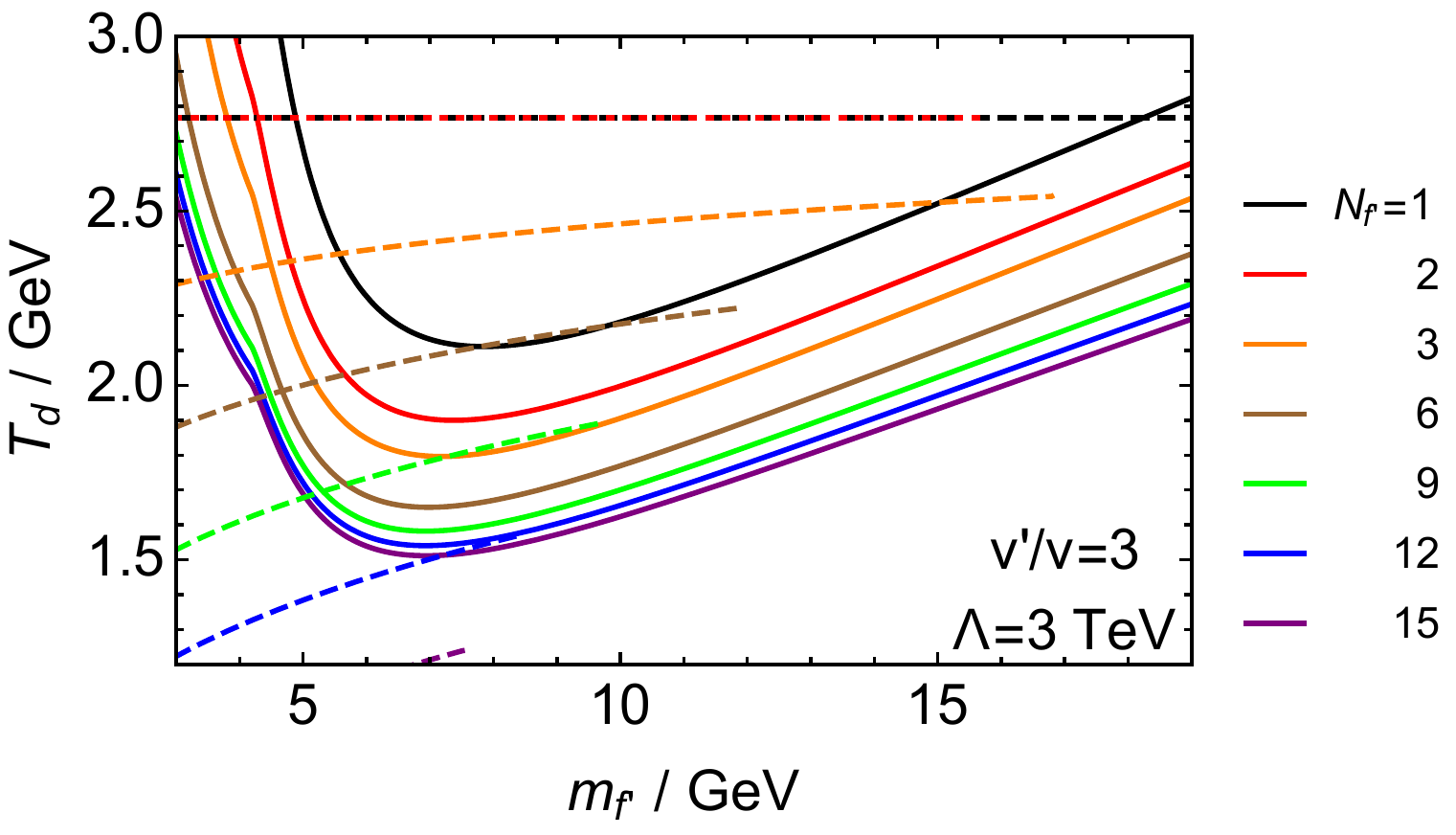}
\includegraphics[clip,width=.45\textwidth]{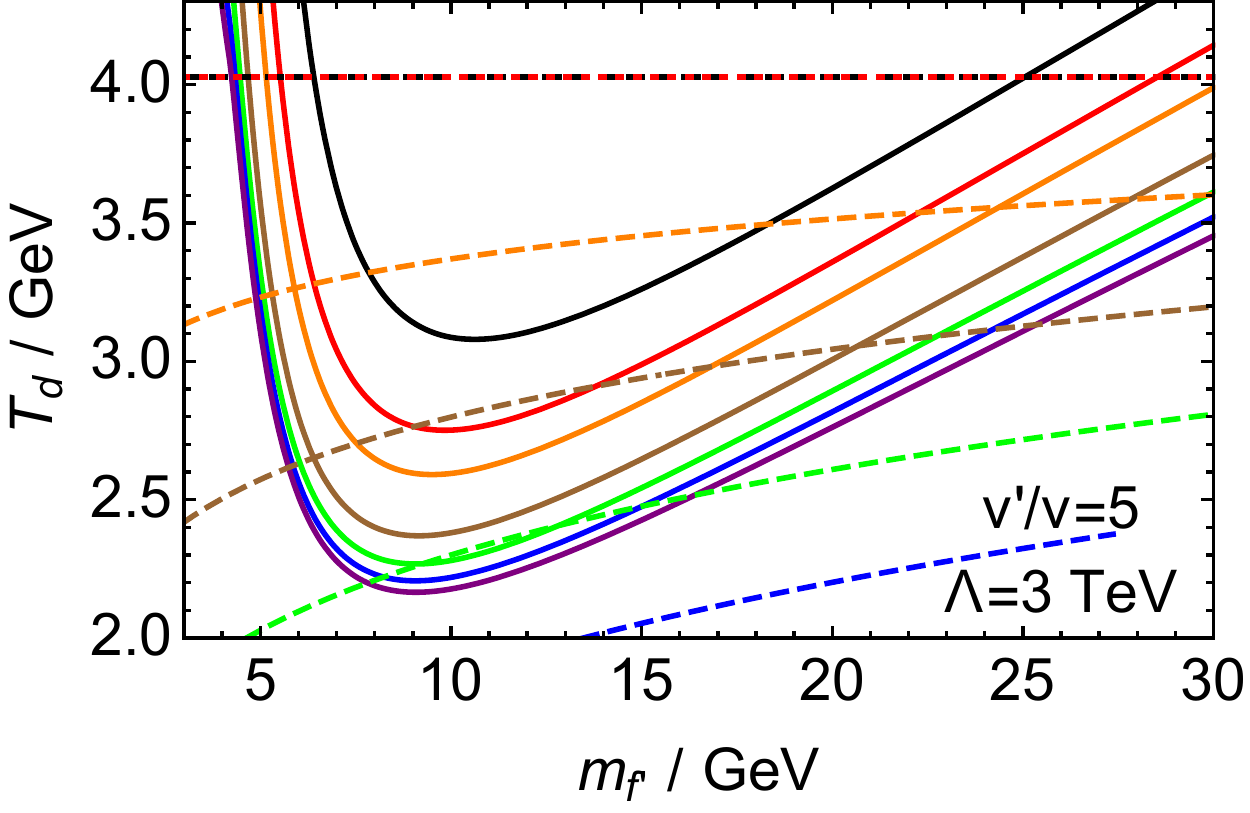}
\caption{
The decoupling temperature, $T_d$, between the two sectors as a function of the mirror fermion mass, $m_{f'}$, for various $N_{f'}$ degenerate states labelled by color.   The dashed lines show the maximal possible mirror QCD phase transition temperature.
}
\label{fig:THiggs}
\end{figure}

In Fig.~\ref{fig:THiggs}, solid curves show the temperature $T_{d}$, below which $({\rm d}\rho' / {\rm d}t)  / \rho' $ is smaller than the expansion rate of the universe, as a function of $m_{f'}$ with various colors for different $N_{f'}=1,2,3,6,9,12,15$.
The black and red dashed lines show the maximal $T_c'$ as estimated in section~\ref{sec:Tc}.
Other dashed lines shows the maximal possible $T_c'$ when $N_{f'}/3$ mirror quarks have a common mass $m_{f'}$, with their color chosen to match those of the corresponding solid lines.
The maximal $T_c'$ are determined in the following way.
We first put as many quarks as possible ($N_+$) above $m_h/2$ with the same mass and determine their mass so that the bound in Eq.~(\ref{eq:bound_yukawa}) is saturated.
We then determine the masses of remaining quarks  ($5-N_+-N_{f'}/3$) so that the bound $\mu >0.75$ is saturated.
(For $v'/v=3,5$ and $\Lambda=3$ TeV, we find that $N_+=5-N_{f'}/3$, and hence the second step is not necessary.)
On the right end of each dashed line, the invisible decay width of the Standard Model Higgs into mirror fermions $f'$ saturate the current bound.
For a mirror fermion mass in the range between abound 5 and 20-30 GeV for $v'/v=3\mathchar`-5$, $T_{d}$ is smaller than $T_c'$, so that the energy of the mirror gluon plasma is efficiently transferred into the Standard Model particles.

\begin{figure}[t]
\centering
\includegraphics[clip,width=.53\textwidth]{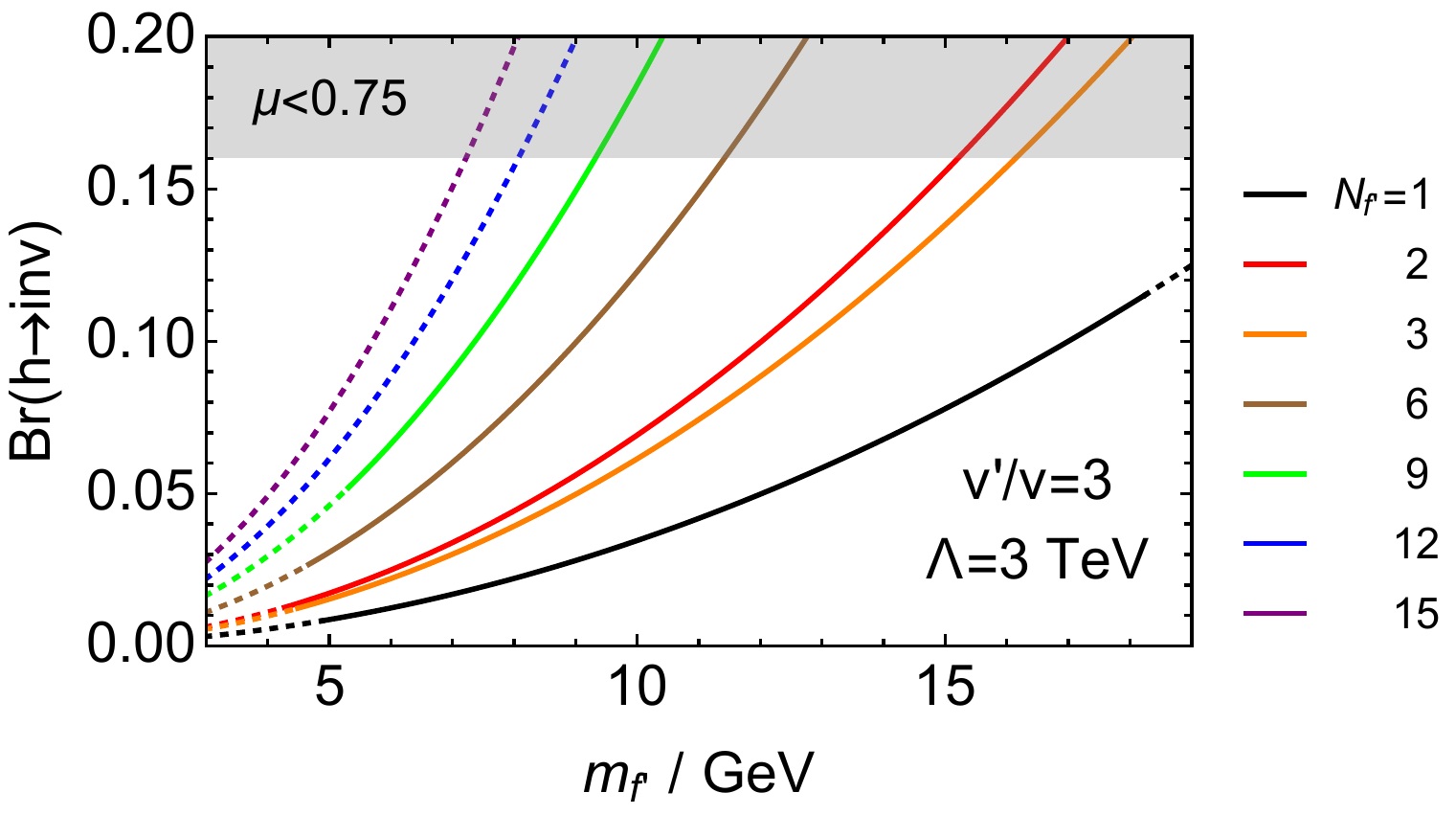}
\includegraphics[clip,width=.43\textwidth]{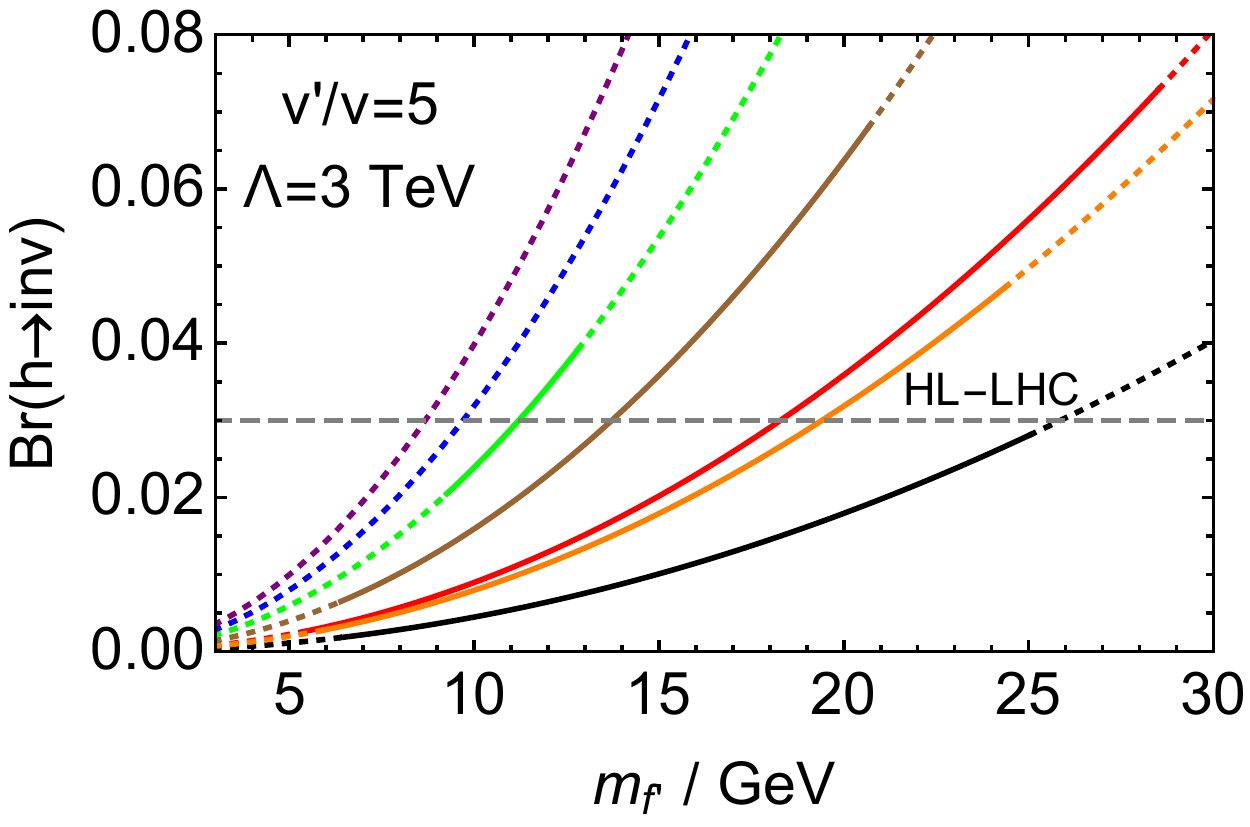}
\caption{
Predictions for the invisible branching ratio of the Standard Model Higgs.
The ranges of $m_{f'}$ that yield $T_d < (>) T_c'$ are depicted by solid (dotted) lines.
}
\label{fig:Brinv}
\end{figure}

In Fig.~\ref{fig:Brinv}, we show the predictions for the invisible branching ratio of the Standard Model Higgs.
The ranges of $m_{f'}$ which yield $T_d > T_c'$ are depicted by dotted lines.
The shaded region is excluded by the measurement of the Higgs-signal strength.
The high-luminosity LHC can probe $\mu<0.93$, whose corresponding invisible branching ratio is shown by a dashed line in the right panel.
It is also sensitive to the invisible branching ratio of 10\%~\cite{ATLAS_prospect}.
For $v'/v\sim 3$, the deviation of the Higgs signal-strength from unity as well as non-zero invisible branching ration may be detected in the high-luminosity running of the LHC.
The ILC is sensitive to an invisible branching fraction of sub-percent level~\cite{Asner:2013psa}.
The region with $T_d<T_c'$ can be probed by the Higgs-signal strength as well as the invisible decay of the Higgs at the high-luminosity LHC and the ILC.


In the above analysis we calculated $T_{d}$
assuming that the dynamics of mirror fermions is described by that of free fermions.
This is not correct for mirror quarks when the temperature is smaller than the binding energy $B_D$ of the mirror QCD interaction.
As the temperature drops below the binding energy,
some mirror quarks form bound states, namely mirror quarkonia.
This effect is expected to enhance the energy transfer rate by the annihilation of mirror fermions.
The annihilation rate of mirror fermions inside quarkonia is
\begin{align}
\Gamma(q'\bar{q}'\rightarrow f\bar{f}) \sim \sigma(q'\bar{q}' \rightarrow f \bar{f})v|_{p_{f'}\simeq m_q' \alpha_3'} \times \frac{1}{4\pi}\left( m_{f'} \alpha_3'\right)^3,
\end{align}
where the second factor is the inverse volume of a quarkonium.
The energy transfer rate by annihilation is given by
\begin{align}
\frac{\rm d}{{\rm d}t}\rho'|_{\rm annihilation} \sim & N_{q'}\sum_f n_B(2m_{q'},T) \Gamma(\eta_{q'}\rightarrow f\bar{f}) \times 2 m_{q'},
\end{align}
where $N_{q'}$ is the number of mirror quarks with a mass $m_{q'}$, and
$n_B(m,T)$ is the number density of a real scalar with a mass $m$ in the thermal bath with a temperature $T$.
The ratio of the energy transfer rate by the annihilation of free fermions to that by the annihilation inside quarkonia is
\begin{align}
\frac{{\rm d}\rho'/{\rm d}t|_{\rm free}}{{\rm d}\rho'/{\rm d}t|_{\rm quarkonia}} \sim \left( \frac{T}{m_{q'} \alpha_3'^2} \right)^{5/2} \sim \left( \frac{T}{E_{\rm B}}\right)^{5/2}.
\end{align}
Here we have used the non-relativistic approximation for $n_{F,B}$.
In the parameter space we have discussed, the binding energy $E_B$ is comparable to the temperature, and hence the formation of the bound state does not change the result in Fig.~\ref{fig:THiggs} significantly.
But we note that it is possible that $T_d<T_c'$ is achieved for a wider parameter region.

Here we show that mirror neutrinos can be in thermal equilibrium down to $T_d$.
Chemical equilibrium of mirror neutrinos is maintained by the annihilation process $f'\bar{f}'\leftrightarrow \nu' \bar{\nu}'$, with cross section
\begin{align}
\sigma(f'\bar{f}'\rightarrow \nu' \bar{\nu}')v = \frac{g_2^4/c_w^4}{64\pi}\left( I_{3f'}-2s_w^2 Q_{f'} \right)^2\frac{m_{f'}^2}{m_{Z'}^{4}}.
\end{align}
The number of mirror neutrinos produced/annihilated per unit volume and time is given by
\begin{align}
\sigma(f'\bar{f}'\rightarrow \nu' \bar{\nu}')v \times n_F(m_{f'},T)^2 \times N_{f'}.
\label{eq:nupfp}
\end{align}
Comparing this rate with $H\times n(\nu')$, we find that chemical equilibrium as well as kinetic equilibrium are maintained down to a temperature of about $m_{f'}/10$.
Kinetic equilibrium alone is maintained by the scattering $f' \nu' \rightarrow f'\nu'$, with a cross section 
\begin{align}
\sigma(f'\nu'\rightarrow f' \nu')v = \frac{g_2^4/c_w^4}{32\pi}\left(\left( I_{3f'}-2s_w^2 Q_{f'} \right)^2 + \frac{1}{4}\right)\frac{T^2}{m_{Z'}^{4}},
\end{align}
and is effective down to a temperature of about $m_{f'}/20$.
Comparing $m_{f'}/10$ and $T_{\rm d}$ in Fig.~\ref{fig:THiggs}, thermal equilibrium of mirror neutrinos is also maintained down to temperature $T_d$.

\subsection{Dark radiation}
\label{sec:DR}

As we have shown, mirror photons and mirror neutrinos can be in thermal equilibrium with Standard Model particles down to temperature $T_d < T_c'$.
Taking account mirror photons and neutrinos, the prediction for $\Delta N_{\rm eff}$ is
\begin{align}
\Delta N_{{\rm eff},\gamma'\nu'} = \frac{4}{7} \times (2 + \frac{7}{4}\times 3) \times \left( \frac{10.75}{g (T_{d})} \right)^{4/3} = 0.29 \times \left( \frac{80}{g (T_{d})} \right)^{4/3},
\label{eq:minimal DR}
\end{align}
which is consistent with the upper bound, $\Delta N_{\rm eff}<0.65$.

\begin{figure}[t]
\centering
\includegraphics[clip,width=.54\textwidth]{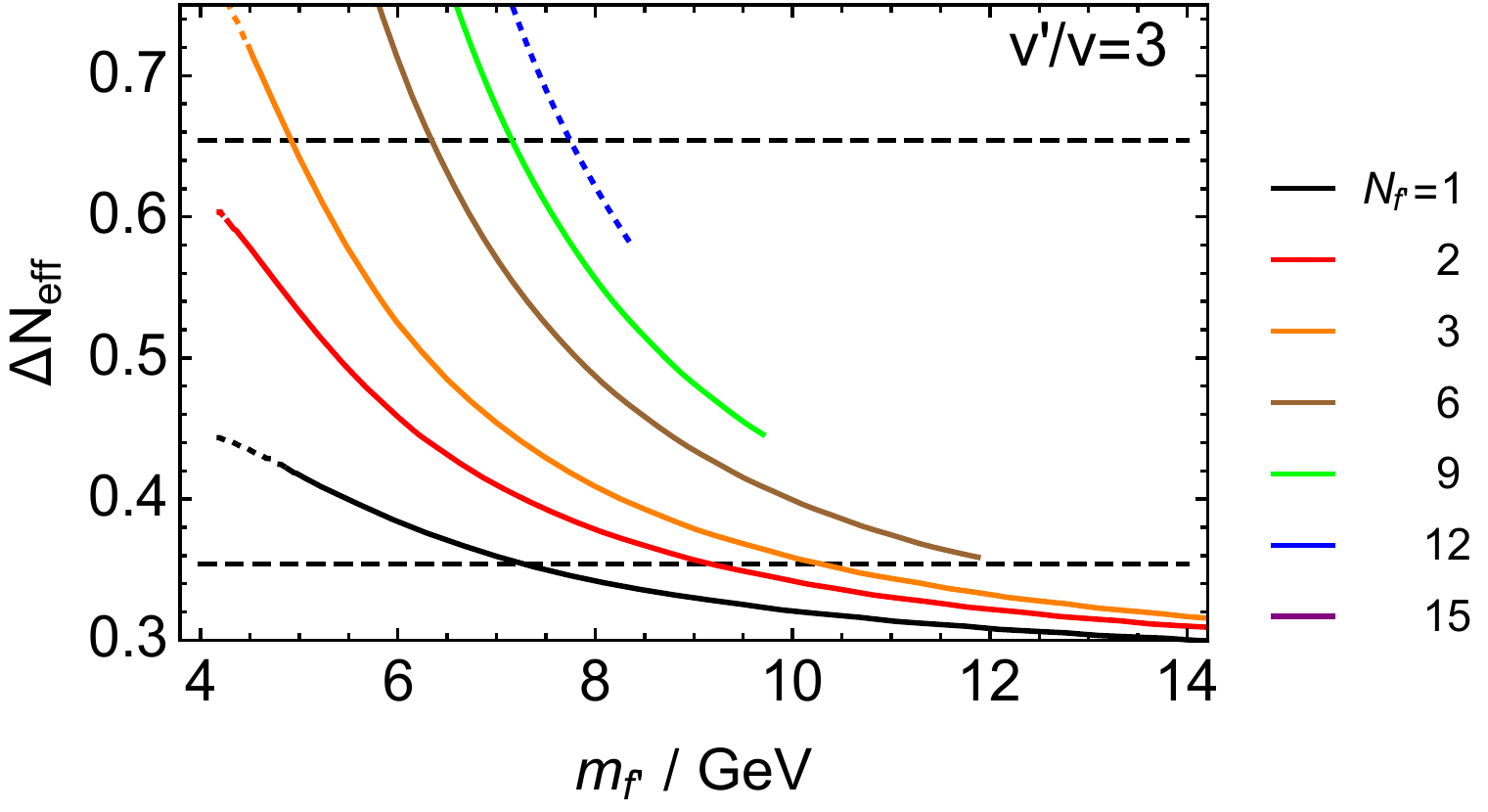}
\includegraphics[clip,width=.45\textwidth]{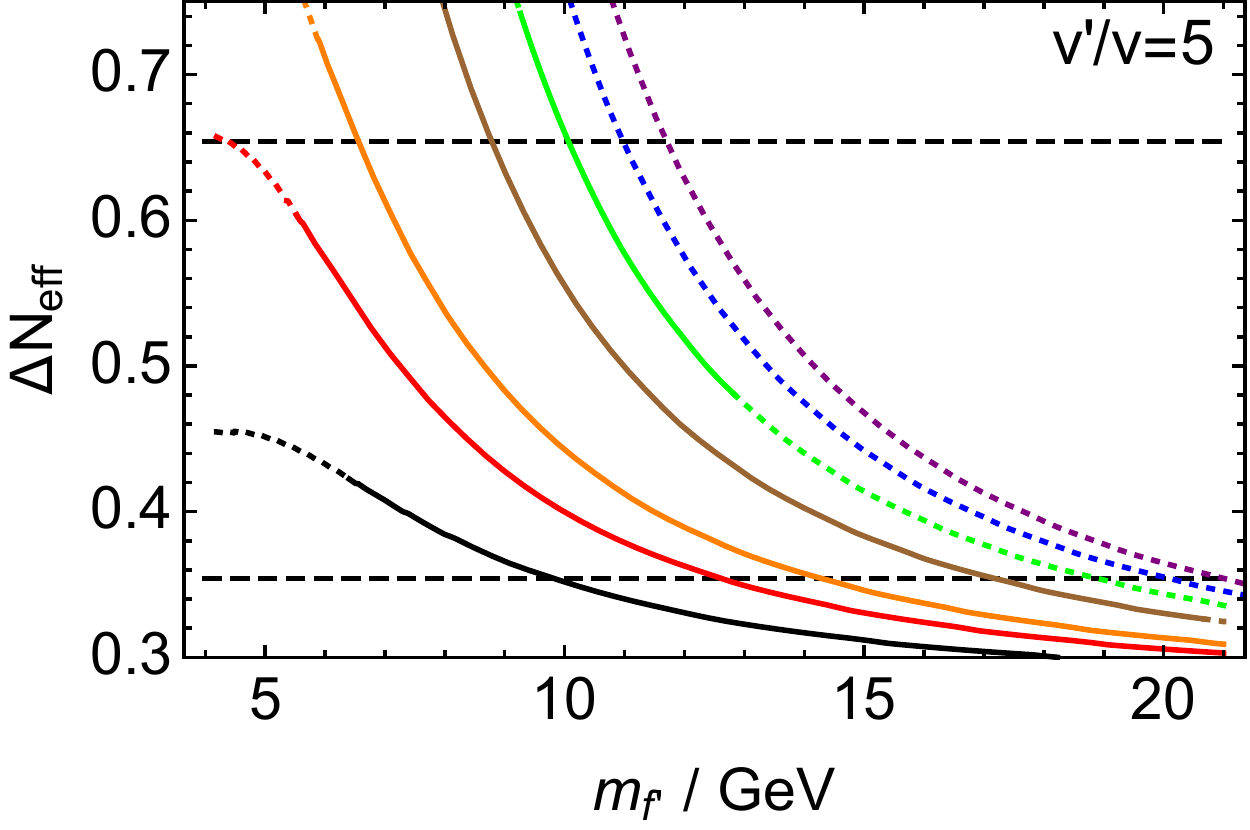}
\caption{
Prediction for $\Delta N_{\rm eff}$ as a function of the mirror fermion mass $m_{f'}$, for various multiplicities, $N_{f'}$. Thermal equilibrium between the two sectors is maintained by the exchange of the SM-like Higgs.  The ranges of $m_{f'}$ that yield $T_d < (>) T_c'$ are depicted by solid (dotted) lines.
}
\label{fig:Neff Higgs}
\end{figure}

To keep $T_d$ smaller than $T_c'$, some mirror fermions must have masses not far above $T_d$ and so they also contribute to dark radiation, giving
\begin{align}
\Delta N_{{\rm eff},\gamma'\nu'f'} = 0.29 \times \left( \frac{80}{g(T_{d})} \right)^{4/3} \left( \frac{7.25 + g_{f}' (T_{d})}{7.25} \right)^{4/3},
\label{eq:DR Higgs}
\end{align}
where $g_{f}'$ is the effective d.o.f.~of mirror fermions $f'$.
In Fig.~\ref{fig:Neff Higgs}, we show the prediction for $\Delta N_{\rm eff}$ as a function of $m_{f'}$.
Here we have neglected the contribution from the mirror gluon plasma, which is correct for $m_{f'}$ that give $T_{d} < T_c'$.
Regions that give $T_{d} > T_c'$ are depicted by dotted lines.
We also assume that $g_{s,f'}$ is well approximated by that of the ideal gas of $f'$.
This is correct for mirror leptons; for mirror quarks, the actual $g_{s,f'}$ is smaller.
Lines in the left panel are terminated if the Higgs coupling-strength falls below 0.75. 
The predicted amount of the dark radiation can be consistent with the experimental bound.

\vspace{0.5cm}
To summarize: for the amount of dark radiation to be below the experimental bound, there must be a mirror fermion with mass in the range $(4-28)$ GeV.

\subsection{Cosmological signals of mirror neutrino masses}

As discussed in section \ref{sec:MinEFT}, the dimension 5 operators of the Minimal Mirror Twin Higgs theory of (\ref{eq:EFT}) result in neutrino masses that are either Majorana, with $m_{\nu'} = (v'/v)^2 m_\nu$, or Dirac with $m_{\nu'} = m_\nu$.   Such masses can have significant effects on structure formation in the universe as well as the CMB spectrum.
Taking into account dilution by Standard Model particles, the $\nu'$ number density is
\begin{align}
n_{\nu'} =  \frac{10.75}{g(T_d)}  \times  \frac{g'(T_d)}{g'_r}  \times n_\nu = \left( \frac{7}{29} \Delta N_{\rm eff}\right)^{3/4}n_{\nu}.
\end{align}

For the case of Majorana neutrinos, the effective total mass of light neutrinos, constrained by data on structure formation, is
\begin{align}
\left(\sum m_\nu \right)_{\rm eff}&\equiv  \sum m_\nu + \sum m_{\nu'} \frac{n_{\nu'}}{n_\nu}  \nonumber \\
 &= \left(\sum m_\nu \right)\times \left( 1 +  \left( \frac{7}{29} \Delta N_{\rm eff}\right)^{3/4} \left( \frac{v'}{v} \right)^2 \right) >2.3 \sum m_\nu.
\end{align}
Here we have used the prediction of Fig.~\ref{fig:Neff Higgs}, $\Delta N_{\rm eff}>0.3$, and the experimental constraint $v'/v>3$, to obtain the last inequality.
Although the current cosmological data are more constraining on $\Delta N_{\rm eff}$ than on $\left(\sum m_\nu \right)_{\rm eff}$,
both parameters may play a comparably important role in observations in the near future.
We note that $\left(\sum m_\nu \right)_{\rm eff}$ can be larger or smaller if the $Z_2$ symmetry is also broken in the Dirac mass term of neutrinos.

For the case of Dirac neutrinos, the effective total mass of neutrinos is
\begin{align}
\left(\sum m_\nu \right)_{\rm eff}
 = \left(\sum m_\nu \right)\times \left( 1 +  \left( \frac{7}{29} \Delta N_{\rm eff}\right)^{3/4}  \right) \simeq \sum m_\nu
\end{align}
so that $m_{\nu'}$ have a small effect on structure formation and the CMB spectrum.

\section{Thermal history including Kinetic mixing}
\label{sec:mixing}

In the last section, we discussed the thermal history of the Minimal Mirror Twin Higgs theory with $\epsilon < 10^{-5}$, so that Higgs exchange is the unique interaction coupling the two sectors, and
found that light mirror fermion must be in the range of about (4-20) GeV.
In this section we allow larger values of $\epsilon$, from the UV completion above $\Lambda_{\rm{TH}}$, so that the Standard Model and mirror sectors also interact 
by kinetic mixing between $U(1)_Y$ and $U(1)_Y'$ gauge bosons, described by the $\epsilon \, B^{\mu \nu} B'_{\mu \nu}$ term of Eq.~(\ref{eq:EFT}).
As we will see, the allowed range of mirror fermions masses are wider than the case without the kinetic mixing.

\subsection{Decoupling temperature}
Here we list the decoupling temperatures of various processes that maintain thermal equilibrium of mirror photons and/or neutrinos.
We take a field basis such that the mirror photon is shifted to eliminate kinetic mixing, $A'\rightarrow A' + \epsilon A $.
In this basis, Standard Model charged particles interact only with photons, while mirror particles interact with both photons and mirror photons.

Mirror photons are in thermal equilibrium with mirror charged fermions $f'$, which
also interact with photons and through mixing, maintaining thermal equilibrium between Standard Model particles and mirror photons. 
The cross section for $f' \gamma' \leftrightarrow f' \gamma$ is given by,
\begin{align}
\sigma (f' \gamma' \leftrightarrow f' \gamma)v = \frac{8\pi}{3} \epsilon^2 \alpha^2 q_{f'}^4 \frac{1}{m_{f'}^2},
\end{align}
where we assume $T\ll m_{f'}$, and $q_{f'}$ is the electromagnetic charge of $f'$.
The scattering rate is smaller than the expansion rate of the universe below a temperature $T_{d,\gamma'}$,
\begin{align}
T_{d,\gamma'} \simeq \frac{m_{f'}}{20 + 2{\rm ln}\frac{\epsilon}{10^{-2}}}.
\end{align}

$U(1)$ kinetic mixing also mixes the mirror $Z'$ boson and the Standard Model photon, allowing
mirror neutrinos to interact with Standard Model fermions with a cross section,
\begin{align}
\sigma(f \nu' \rightarrow f \nu') v \simeq \sigma(f \bar{f} \leftrightarrow \nu' \bar{\nu}') v \simeq \frac{16\pi}{3} q_f^2 \frac{\alpha^2}{c_w^4} \epsilon^2 \frac{T^2}{m_{Z'}^{4}}.
\end{align}
This scattering becomes ineffective below the temperature $T_{d,\nu'}$,
\begin{align}
T_{d,\nu'}\simeq 0.3~{\rm GeV} \left( \frac{\epsilon}{10^{-2}}\right)^{-2/3} \left( \frac{v'/v}{3}\right)^{4/3}.
\end{align}
$T_{d,\nu'}$ must be larger than the QCD phase transition temperature, otherwise the mirror neutrino abundance exceeds the upper bound on dark radiation. This gives an upper bound on kinetic mixing, $\epsilon \lesssim 10^{-2}$,
so that $T_{d,\gamma'}\gtrsim m_{f'}/20$.

Mirror photons are in the thermal equilibrium with mirror charged fermons $f'$.
If the scattering rate of the process $f'\bar{f}' \leftrightarrow \nu'\bar{\nu}'$ is large enough, kinetic as well as chemical equilibrium of mirror neutrinos is maintained. As we have seen, this interaction is effective down to the temperature of $m_{f'}/10\equiv T_{d,\nu'\gamma',{\rm che}}$.
Kinetic equilibrium of mirror neutrinos is maintained by $f'\nu'\rightarrow f'\nu'$,
which is efficient down to a temperature of $m_{f'}/20 \equiv T_{d,\nu'\gamma',{\rm kin}}$.

\subsection{Dark radiation}

From the above consideration on decoupling temperatures, we obtain a bound on mirror fermion masses.
In order for the energy of mirror gluons to be transferred into Standard Model particles, $T_{\gamma'}$ or max$(T_{\nu'}, T_{d,\nu'\gamma',{\rm kin}})$ must be smaller than $T_c'$. This requires a charged mirror fermion lighter than $20T_c'=20\mathchar`-50$ GeV.
Charged mirror fermion masses are also bounded from below.
Even if mirror and Standard Model sectors decouple just before the QCD phase transition, $\Delta N_{\rm eff}$ is too large unless charged mirror leptons have masses larger than 2 GeV.
Since mirror quarks are bound to form mesons below $T_c'$, they can be lighter than 2 GeV as long as the meson masses are above 2 GeV.
The allowed range of mirror fermion masses is wider than the case without the kinetic mixing.

\begin{figure}[t]
\centering
\includegraphics[clip,width=.55\textwidth]{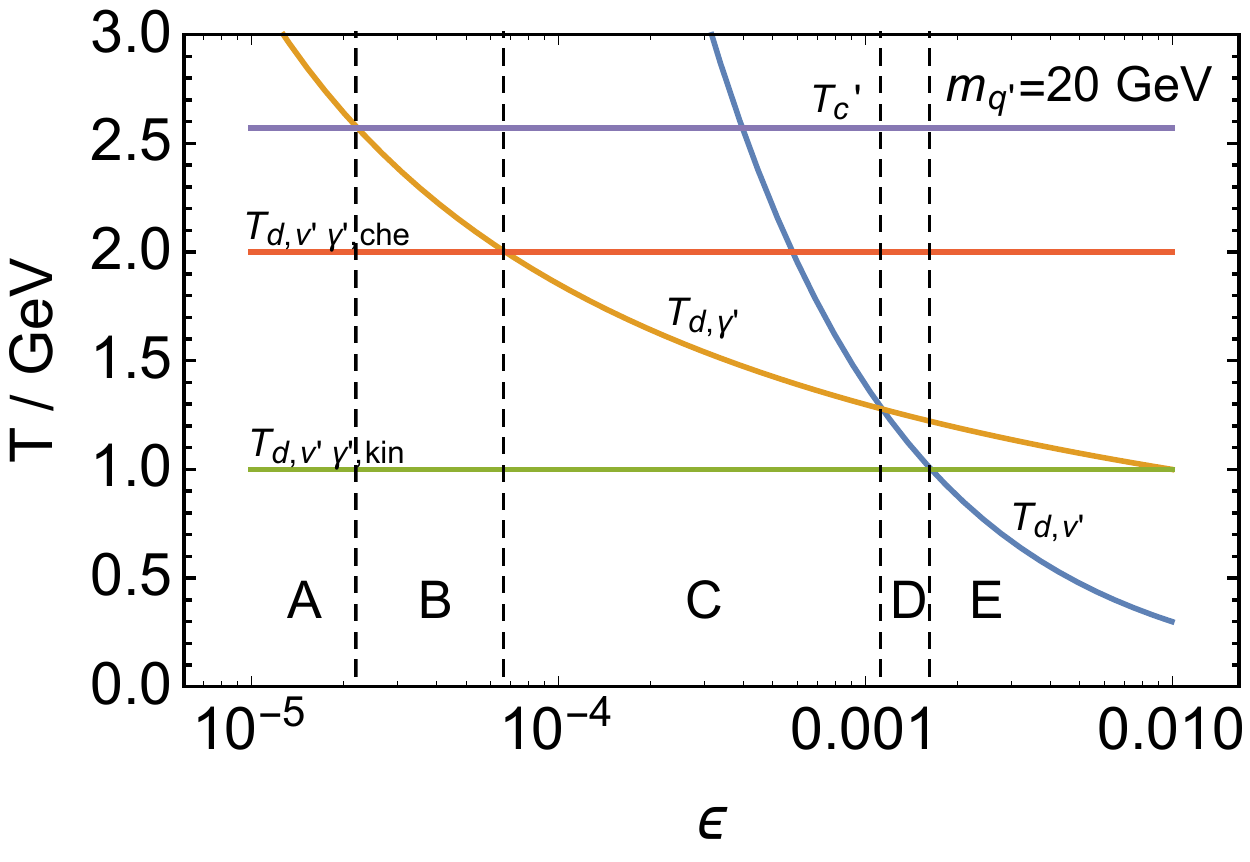}
\caption{
Decoupling temperatures of various process when the lightest mirror charged fermion is a quark of mass 20 GeV.
}
\label{fig:temps}
\end{figure}

Here we estimate $\Delta N_{\rm eff}$ for a representative point of parameter space, illustrating the importance of a variety of
reactions between mirror and QCD phase transitions.
Suppose that the lightest mirror charged fermion is a quark of mass 20 GeV, so that
mirror glueballs decay into mirror photons just below $T_c'$.
In Fig.~\ref{fig:temps}, we show the decoupling temperatures of various processes as a function of $\epsilon$. We also show the mirror QCD phase transition temperature, which we assume to be the maximal one we estimated in section~\ref{sec:Tc} for $v'/v=3$ and $\Lambda = 3$ TeV.
In region A, kinetic mixing is insufficient to transfer the energy of the mirror gluon plasma into Standard Model particles, and $\Delta N_{\rm eff}$ is determined by Higgs exchange.
In region B $\Delta N_{\rm eff}$ is given by that in Eq.~(\ref{eq:minimal DR}) with $T_d = T_{d,\gamma'}$.
In region C, the number density of mirror neutrinos per comoving volume is conserved below $min(T_{d,\nu'\gamma',_{\rm che}},T_{d,\nu'})\equiv T_{d,\nu',{\rm che}}$, giving
\begin{align}
\Delta N_{\rm eff} \simeq \frac{4}{7} \times 2 \; \left( \frac{10.75}{g (T_{d,\gamma'})} \right)^{4/3} + 3 \; \left( \frac{10.75}{g (T_{d,\gamma'})} \right)^{4/3}
\left(
1 - \frac{4860 \zeta(3)^2}{7\pi^6} \frac{g(T_{d,\nu',{\rm che}}) -g(T_{d,\gamma'}) }{g(T_{d,\nu',{\rm che}}) + 2}
\right).
\end{align}
The last factor in the second term accounts for the conservation of the comoving number density of mirror neutrinos.
However, we find this factor is $\geq0.94$, so that $\Delta N_{\rm eff}$ is well approximated by Eq.~(\ref{eq:minimal DR}) with $T_d = T_{d,\gamma'}$.
In region D,  $\Delta N_{\rm eff}$ is given by Eq.~(\ref{eq:minimal DR}) with $T_d = T_{d,\nu'}$.
In region E, mirror photons decouple from the thermal bath at $T_{d,\nu'\gamma',{\rm che}}$,
while mirror neutrinos decouple at $T_{d,\nu'}$, giving
\begin{align}
\Delta N_{\rm eff} = \frac{4}{7} \times 2 \; \left( \frac{10.75}{g (T_{d,\nu'})} \right)^{4/3} \left( \frac{ g (T_{d,\nu'}) + 5.25 }{ g (T_{d,\nu'\gamma',{\rm kin}}) + 5.25 } \right)^{4/3} + 3 \;  \left( \frac{10.75}{g (T_{d,\nu'})} \right)^{4/3}.
\end{align}
We find that in region B, C, and D, $\Delta N_{\rm eff}$ is about 0.3. In region E, $\Delta N_{\rm eff}$ is larger than 0.3, and can saturate the bound on $\Delta N_{\rm eff}$ for $\epsilon\simeq 10^{-2}$.

If the lightest mirror quark is heavier than 20 GeV, mirror gluons do not decay immediately below $T_c'$, but decay later.
This is excluded if kinetic mixing is absent, because $T_c'$ and $T_d$ are close to each other.
With kinetic mixing, the energy of mirror photons can be transferred into SM particles well below $T_c'$, if a charged mirror lepton is light enough.
As long as mirror glueballs decay into mirror photons before the QCD phase transition, sufficiently light mirror charged leptons can suppress $\Delta N_{\rm eff}$ to be within the allowed range.

\subsection{Milli-charged particle}
With $U(1)$ kinetic mixing, mirror fermions of mirror charge $q_{f'}$ can be understood to carry SM electric charge $\epsilon q_{f'}$.
In the range $m_{f'}=O(1\mathchar`-10)$ GeV,
the most stringent constraint comes from collider experiments~\cite{Davidson:1991si,Jaeckel:2012yz}, which is much weaker than the bound from $\Delta N_{\rm eff}$, $\epsilon\lesssim10^{-2}$.
A proposed search at the LHC can search down to mixings of $\epsilon = 10^{-2}-10^{-3}$~\cite{Haas:2014dda}.

\section{Mirror baryon dark matter}
\label{sec:DM}
The mirror sector, like the SM, possesses accidental baryon and lepton symmetries.
Mirror baryons and leptons may
account for dark matter in the universe~\cite{Goldberg:1986nk} (see~\cite{Blinnikov:1982eh,Kolb:1985bf} for earlier work on astrophysical considerations of mirror baryons and leptons.).
We assume a non-zero mirror matter-antimatter asymmetry with the asymmetric component comprising dark matter.
The proximity of the energy densities of baryons and dark matter could be understood
if the sectors which generate the Standard Model and mirror asymmetry are close to $Z_2$ symmetric, and the dark matter mass is $O(1\mathchar`- 30)$ GeV.

We consider the mirror up quark $u'$, down quark $d'$ and electron $e'$ as possible components of dark matter.
To simplify the discussion, and motivated by section IVC, we take the masses of $u'$ and $d'$ larger than the dynamical scale of mirror QCD, so that the masses of mirror baryons mainly originate from mirror quark masses.

\subsection{Dark matter candidates}
We consider dark matter candidates in Minimal Mirror Twin Higgs, where the $Z_2$ symmetry is broken solely by Yukawa couplings.
A dark matter candidate depends on the mass spectrum of $u'$, $d'$ and $e'$. Below we consider several interesting possibilities.

\subsubsection{Light $u'$, $d'$: mirror neutron}
\label{sec:udnu}
Taking $e'$ sufficiently heavy, as the temperature drops below $m_{e'}$, it decays into $u'$, $d'$ and $\nu'$ and its abundance becomes negligible.  
For the remaining $u'$, $d'$, number changing processes from $W'$ exchanges are absent,
so that $u'$ and $d'$ are both stable and have separately conserved comoving numbers.  Mirror charge neutrality implies $n_{d'} = 2 n_{u'}$.  Below the mirror QCD phase transition temperature, $u'$ and $d'$ are combined into mirror baryons $B'$ and mesons $M'$. The stable hadrons are
\begin{align}
B'_{uuu},~B'_{uud},~B'_{udd},~B'_{ddd},~M'_{u\bar{d}},\nonumber
\end{align}
with an obvious notation.
The meson $M'_{u\bar{d}}$ is captured by mirror baryons, e.g.~
\begin{align}
M'_{u\bar{d}} + B'_{ddd}\rightarrow B'_{udd}~(+ \gamma').\nonumber
\end{align}
As the sum of the masses of $M'_{u\bar{d}}$ and $B'_{ddd}$ is larger than the mass of $B'_{udd}$ by $2 m_{d'}$, the meson $M'_{u\bar{d}}$ disappear from the thermal bath by this capture process.
Finally, $B'$ scatter with each other and almost all become the mirror neutron $B'_{udd}$. For example, the scattering process
\begin{align}
B'_{uud} + B'_{ddd}\rightarrow 2 B'_{udd}~(+ \gamma')\nonumber
\end{align}
eliminates $B'_{uud}$ and $B'_{ddd}$ from the thermal bath.
Note that the sum of the masses of $B'_{uud}$ and $B'_{ddd}$ is larger than twice of the mass of $B'_{udd}$ as the mirror baryon $B'_{ddd}$ is spin-$3/2$ and has a contribution to its mass from a spin-spin interaction.

Without any IR effects, the scattering cross section between mirror neutrons is $O(m_{B'_{udd}}^{-2})$, and does not affect structure formation.
The cross section can be enhanced up to the unitarity limit by some IR effects, e.g.~the Sommerfeld effect~\cite{Sommerfeld,Hisano:2002fk} or the existence of resonance states~\cite{Cline:2013zca}.
In our case, mirror pions are also heavy due to large mirror quark masses, and those effects are expected to be suppressed.

\subsubsection{Light $d'$, $e'$: mirror atom}
\label{sec:denu}
If $u'$ is sufficiently heavy, it decays to $d'$, $e'$ and $\nu'$ and disappears from the thermal bath. The mirror electron $e'$  is now stable, as its decay is kinematically forbidden.
After the QCD$'$ phase transition, $d'$ combines into $B'_{ddd}$.
To preserve charge neutrality, there are the same number of $\bar{e}'$.
The mirror baryons $B'_{ddd}$ and positrons $\bar{e}'$ eventually ``recombine" into mirror atoms.

The self scattering cross section of mirror atoms is given by~\cite{Cline:2013pca}
\begin{align}
\sigma/ {m_D} \simeq \frac{100}{\alpha^2} \frac{(R+1)^4}{R^2} \frac{1}{m_D^3} = 6.6 \, {\rm cm}^2/{\rm g} \times \left(\frac{10~{\rm GeV}}{m_D}\right)^3 \frac{(R+1)^4}{16R^2},
\end{align}
where $R = {\rm max} (m_{B'_{ddd}}/m_{e'},m_{e'}/m_{B'_{ddd}})$ and $m_D$ is the mass of the dark atom.
Here we assume that $R\sim 1$ and the kinetic energy of the dark atom is much smaller than the binding energy.
The cross section is minimized for $R=1$, which we assume in the following.
The cored dark matter halo profiles could be explained by $m_D\sim 10$ GeV.
The upper bound on the cross section in dwarf galaxies, $\sigma/m_D < 10 \, {\rm cm}^2/{\rm g}$~\cite{Kaplinghat:2015aga} requires that $m_D>9~{\rm GeV}$.
The constraint from galaxy clusters is weak, since the velocity dispersion of dark matter is so large that  its kinetic energy is comparable to the binding energy, and the self interaction cross section is suppressed.

The recombination of mirror baryons and electrons is incomplete. From the numerical estimation in Ref.~\cite{CyrRacine:2012fz} we obtain the ionized fraction,
\begin{align}
x_{e'} \simeq 0.02 \times \left( \frac{m_D}{10~{\rm GeV}} \right)^2  \frac{4R}{(R+1)^2}.
\end{align}
The ionized components interact with each other and with atomic dark matter strongly, and may affect halo shape~\cite{CyrRacine:2012fz} and the scattering of clusters.
The estimation of these and other constraints on the ionized fraction is beyond the scope of this paper.

With the sizable kinetic mixing that we considered in section~\ref{sec:mixing},
the ionized components participate in acoustic oscillations and affect the CMB spectrum~\cite{Dubovsky:2001tr,Dubovsky:2003yn}%
\footnote{%
\small Mirror baryons and electrons interact with SM particles via kinetic mixing and are heated, which may change the ionized fraction.
}.
The upper bound on the ionized fraction is
$\Omega_{\rm ionized}/\Omega_{\rm atom}<0.01$~\cite{Dolgov:2013una}
which, together with the limit from self scattering in dwarfs, disfavors atomic dark matter for such kinetic mixing%
\footnote{%
\small The galactic magnetic field prevents the ionized component from entering the disk,
and ionized components initially in the disk are likely to escape the disk owing to Fermi acceleration by Super-Nova remnants.  Furthermore, the magnetic field of the Earth prevents any ionized component from reaching the Earth.  Hence, the ionized component may evade direct detection experiments performed on the Earth~\cite{Chuzhoy:2008zy}.

}.

\subsubsection{Light $u'$, $e'$: mirror atom}
\label{sec:uenu}
If $d'$ is sufficiently heavy, $B'_{uuu}$ and $e'$ are stable and eventually form a mirror atom, $(B'_{uuu} + 2e')$
with a binding energy larger than that of $(B'_{ddd}$ + $\bar{e}')$. The constraints from self interactions and the ionized components of the atomic dark matter are weakened.
The precise determination of the constraints is beyond the scope of this paper.

\subsubsection{Light $u'$, $d'$, $e'$: mirror neutron}
Let us assume that $u'$, $d'$, $e'$ are all light, with a spectrum such that $e'$, $u'$, $d'$ are stable, $m_{u'} + m_{d'} > m_{e'}$, $m_{e'} + m_{d'} > m_{u'}$, $m_{e'} + m_{u'} > m_{d'}$. (Other mass spectra belong to one of the fore-mentioned three cases.) Then the following mirror particles are stable,
\begin{align}
B'_{uuu},~B'_{uud},~B'_{udd},~B'_{ddd},~e'.\nonumber
\end{align}
With $u'$, $d'$, and $e'$ all light, reactions mediated by $W'$ result in the removal of the $e'$, for example by $e' u' \rightarrow \nu' d'$.  
Scattering among the various $B'$ remove $B'_{uuu}$ and $B'_{ddd}$ and mirror charge neutrality implies that only the mirror neutron remains.

In the expanding universe, however, the $W'$ mediated interaction may freeze out before mirror protons and mirror electrons are removed from the thermal bath.
At $T\lesssim m'$, where $m'$ is the mass scale of mirror electrons and mirror protons,
the freeze-out temperature is given by solving
\begin{align}
\frac{g_2^4}{8\pi}\frac{m^{'2}}{m_{W'}^{4}}\times \frac{\rho_{\rm DM}/s}{m'}s(T) = H(T),
\end{align}
where $s(T)$ is the entropy density, and $\rho_{\rm DM}/s \simeq 3.6\times 10^{-9} {\rm GeV}$ is the energy density of dark matter divided by the entropy density.
The first factor in the left-hand side is the cross section of the $W'$ mediated process, and the second factor is a rough estimate on the number density of mirror protons and electrons.
The freeze-out temperature is then given by
\begin{align}
T_{{\rm fo},W'}\simeq 1~{\rm GeV}\times \frac{5~{\rm GeV}}{m'} \left(\frac{v'/v}{3}\right)^4.
\end{align}
If the freeze-out temperature is much smaller than $m_p'+m_e'-m_n'$, mirror protons and mirror electrons disappear from the thermal bath. This is the case for $m'\gtrsim 10$ GeV.
For $m'\lesssim 10$ GeV, a non-negligible amount of mirror electrons and protons remain.
Some of them later recombine into mirror atomic states.
Unlike the cases of sections~\ref{sec:denu} and \ref{sec:uenu}, these states decay into mirror neutrons and mirror neutrinos through mirror electron capture.
Thus there is no constraint from the self-interaction of atomic dark matter.
However, there is a constraint on the ionized component.

\subsection{Direct detection of dark matter}

The above dark matter candidates, $N'$, interact with Standard Model nucleons, $N$, through the exchange of the Standard Model Higgs, and may be observed in direct detection experiments~\cite{Craig:2015xla,Garcia:2015toa,Farina:2015uea}.
The interaction of the Standard Model Higgs relevant for the direct detection is given by
\begin{align}
{\cal L} &=
\frac{h}{\sqrt{2}v} \left[- \left(\frac{v }{v'}\right)^2 \sum_{f'\in N'}m_{f'}f' \bar{f}' -  \sum_{q = u,d,s}m_{q}q \bar{q} + \sum_{q=c,b,t}\frac{\alpha_3}{12\pi} \left( 1 +  \frac{11}{4}  \frac{\alpha_3(m_q)}{\pi} \right)  G_{\mu\nu}^a G^{a\mu\nu} \right] \nonumber \\
&=\frac{h}{\sqrt{2}v} \left[- \left(\frac{v }{v'}\right)^2 \sum_{f'\in N'}m_{f'}f' \bar{f}' -  \sum_{q = u,d,s}m_{q}q \bar{q} + 3.5\times \frac{\alpha_3}{12\pi}G_{\mu\nu}^a G^{a\mu\nu} \right],
\end{align}
where we have taken into accout the one-loop QCD correction to the coupling with gluons~\cite{Djouadi:2000ck}.
The relevant matrix elements of $N'$ is given by
\begin{align}
\sum_{f'\in N'}<N'|m_{f'}f' \bar{f}'|N'> = 2 m_{N'}^2.
\end{align}
Here we assume that the mass of $N'$ is mainly given by the masses of mirror fermions.
Using the trace anomaly formula for the Standard Model nucleon~\cite{Shifman:1978zn},
\begin{align}
2 m_N^2 = <N| T^\mu_\mu  |N> &= - \frac{9 \alpha_3}{8\pi} <N| G_{\mu\nu}^a G^{a\mu\nu} |N> + \sum_{q=u,d,s}<N| m_q q \bar{q}   |N>,
\end{align}
and matrix elements derived by a lattice calculation~\cite{Abdel-Rehim:2016won}%
\footnote{%
\small This matrix element, obtained from a lattice calculation, is consistent with the one extracted from hadron scattering data with the aid of chiral perturbation theory~\cite{Alarcon:2011zs,Alarcon:2012nr}.
},
\begin{align}
\sum_{q=u,d,s}<N| m_q q \bar{q}   |N> \simeq 0.1 \times 2 m_N^2,
\end{align}
the scattering cross section between $N'$ and $N$ through Higgs exchange is given by
\begin{align}
\sigma_{NN'} = \frac{0.028}{\pi} \frac{m_{N'}^2 m_N^2}{v'^4 m_h^4} \left( \frac{m_N m_{N'}}{m_N + m_{N'}} \right)^2.
\end{align}

If kinetic mixing is absent, $m_{N'}$ is bounded from below.
Let us consider the mirror neutron, which is free of constraints from self interactions as well as from the efficiency of mirror recombination.
It is made of one $u'$ and two $d'$ and
the lines  in Fig.~\ref{fig:Neff Higgs} for $N_{f'}=6$ show that
the constraint on $\Delta N_{\rm eff}$ requires 
the mirror neutron to be heavier than $19$ GeV ($26$ GeV) at the $2\sigma$ level and $36$ GeV ($51$ GeV) at the $1\sigma$ level for $v'/v=3$ ($5$).
The mirror neutron mass of $m_{N'}= O(1\mathchar`-30)$ GeV can explain the proximity of the energy densities of baryons and dark matter. Combined with the constraint from $\Delta N_{\rm eff}$, $m_{N'}=O(10)$ GeV is an interesting region.
Note that, however, $\Delta N_{\rm eff}$ is estimated by the ideal gas approximation of mirror quarks. The actual lower bound on $m_{N'}$ can be weaker.

\begin{figure}[t]
\centering
\includegraphics[clip,width=.55\textwidth]{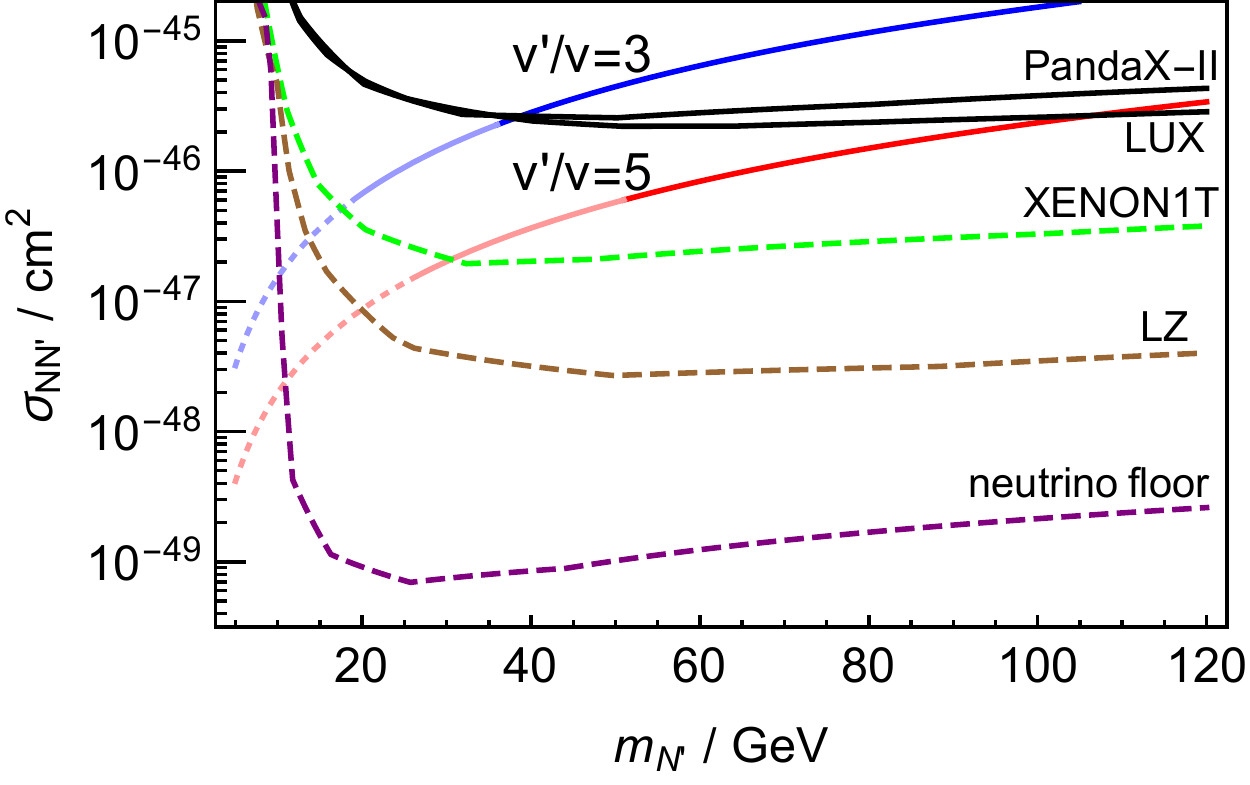}
\caption{
The scattering cross section between dark matter $N'$ and a Standard Model nucleon $N$ as a function of the mass of dark matter.
Here we assume that the mass of dark matter is given by mirror fermion masses, and the matrix element of the trace anomaly is saturated by the mirror fermion mass terms.
}
\label{fig:NNp scat}
\end{figure}

In Fig.~\ref{fig:NNp scat}, we show the prediction on $\sigma_{NN'}$ as a function of $m_N'$.
The regions depicted by thin (dotted) lines are disfavored by $\Delta N_{\rm eff}$ at the $1 (2) \sigma$ level,
if the mirror neutron is dark matter and kinetic mixing is absent.
We show the constraints from the LUX experiment~\cite{Akerib:2016vxi} and the Panda-II experiment~\cite{Tan:2016zwf} by solid lines.
The higher mass region, $m_{N'}>40~(110)$ GeV is excluded for $v'/v=3~(5)$.
We also show the sensitivity of the XENON1T experiment~\cite{Aprile:2015uzo}, the LZ experiment~\cite{Akerib:2015cja}, and the neutrino floor~\cite{Billard:2013qya} by dashed lines.
We conclude that dark matter can be detected by near future experiments in the region
consistent with the upper bound on $\Delta N_{\rm eff}$ without kinetic mixing.

Once kinetic mixing between hypercharge gauge bosons is introduced, $m_N'$ may be smaller and hence direct detection via Higgs exchange may be difficult to observe.
In this case, however,
direct detection via the kinetix mixing is possible.
The mirror neutron dark matter interacts 
through its magnetic dipole moment.
The magnetic moment of mirror neutron is expected to be of order
\begin{align}
\mu_{N'} \sim \epsilon e \frac{1}{m_{N'}}.
\end{align}
Translating the bound derived in~\cite{DelNobile:2014eta}, we obtain the bound $\epsilon\lesssim 10^{-3}\mathchar`-10^{-4}$
for $m_{N'}=O(1\mathchar`-10)$ GeV.
Projected low-threshold experiments such as CRESST III~\cite{Angloher:2015eza} and SuperCDMS SNOLAB~\cite{Calkins:2016pnm} will probe smaller kinetic mixing.

Mirror atoms have a large radius and interact through screened charges of mirror baryons and electrons.
The scattering cross section between atomic dark matter and a nucleus (per nucleon) is given by~\cite{Cline:2012is}
\begin{align}
\sigma \simeq \frac{4\pi \alpha^2 \epsilon^2}{\alpha_{\rm bind}^4m_{N'}^2} \simeq 10^{-34}~{\rm cm}^2 \left( \frac{\alpha}{\alpha_{\rm bind}} \right)^4 \left( \frac{\epsilon}{10^{-5}} \right)^2 \left(\frac{10~{\rm GeV}}{m_{N'}}\right)^2 ,
\end{align}
where $\alpha_{\rm bind}$ is the fine-structure constant of the binding force that forms the atom.
Here we assume $R\simeq 1$.
A kinetic mixing that allows $T_d <T_c'$ ($\epsilon>10^{-5}$) is excluded by various direct detection experiments.
Especially, the light mass region ($m_{N'}=O(1)$ GeV) is excluded by CRESST II~\cite{Angloher:2015ewa}).

\section{Non-Minimal $Z_2$ Breaking From $B-L$ vevs}
\label{sec:heavynu'}

In this section we consider $Z_2$ symmetry breaking beyond Minimal Mirror Twin Higgs.
While the SM neutrinos become very light via the seesaw mechanism,
mirror neutrinos can be much heavier if the mirror right-handed neutrinos do not obtain large masses.
This can be achieved if the $Z_2$ symmetry is also spontaneously broken by a large VEV breaking $B-L$ symmetry and a small/vanishing VEV breaking $B'-L'$.

\subsection{Dark radiation}

Let us first consider the case where a $B'-L'$ gauge field is absent, or the gauged $B'-L'$ symmetry is broken at an intermediate mass scale and hence the $B'-L'$ gauge field is heavy%
\footnote{\small In the supersymmetric Twin Higgs model, this is naturally achieved by a condensation of a mirror right-handed sneutrino at the TeV scale.}.
The prediction on $\Delta N_{\rm eff}$ in Eq.~(\ref{eq:DR Higgs}) becomes
\begin{align}
\Delta N_{{\rm eff},\gamma' f'} \; =\; & \frac{4}{7}
\times 2 \; \left( \frac{10.75}{g(T_{d})}\right)^{4/3}\left( \frac{2 + g_{f}' (T_{d})}{2} \right)^{4/3}
\;=\;  0.08 \; \left( \frac{80}{g (T_{d})} \right)^{4/3} \left( \frac{2 + g_{f}' (T_{d})}{2} \right)^{4/3}.
\label{eq:DR heavy nu}
\end{align}
$\Delta N_{\rm eff}$ can be as small as $0.08$.

Next we consider the case with a massless $B'-L'$ gauge field.
As we will see in the next sub-section, the massless $B'-L'$ gauge field is beneficial in identifying mirror baryons with dark matter.
The prediction for  the amount of the dark radiation are given by
\begin{align}
\Delta N_{{\rm eff},\gamma' f'} \;=\; \frac{4}{7} \times 4
\; \left( \frac{10.75}{g_{*s} (T_{d})}\right)^{4/3} \left( \frac{4 + g_{f}' (T_{d})}{4} \right)^{4/3}
\;=\; 0.17 \; \left( \frac{80}{g (T_{d})} \right)^{4/3} \left( \frac{4 + g_{f}' (T_{d})}{4} \right)^{4/3}.
\label{eq:DR heavy nu2}
\end{align}
$\Delta N_{\rm eff}$ can be as small as $0.17$.

\subsection{Additional dark matter candidates}
The mirror neutrino masses now arise from Dirac Yukawa couplings  and the lightest, $\nu'$, could have a mass of order the lighter states of $u'$, $d'$ and $e'$, or could be heavier than these light states.  Also, a massless $B'-L'$ gauge field may alter dark matter phenomenology.
When $B'-L'$ is exact, so that $B'-L'$ charge is conserved, the $B'$ and $L'$ asymmetries must be produced after the mirror sphaleron process freezes out to avoid washout of $B'$ and $L'$ asymmetries.
On the other hand,  if the $B'-L'$ symmetry is only approximate,
a $B'-L'$ asymmetry may be produced before the mirror sphaleron process freezes out.

\subsubsection{Light $u'$, $d'$, $\nu'$ with unbroken $B'-L'$ gauge symmetry: mirror neutron and neutrino}
Let us assume an unbroken $B'-L'$ gauge symmetry in the setup of section~\ref{sec:udnu}.
Then the mirror neutron itself is no longer a dark matter candidate, as it feels a long-range force by the unbroken $B'-L'$ gauge field. However, a mirror neutron and a mirror neutrino, whose number densities are identical due to $B'-L'$ charge neutrality, can form an atom. The bound from self interactions as well as the ionized fraction is evaded for $\alpha_{B-L}' \gtrsim 10^{-2}$.

\subsubsection{Light $d'$, $e'$, $\nu'$ with unbroken $B'-L'$ gauge interaction: mirror atom}
With unbroken $B'-L'$ gauge symmetry in the setup in section~\ref{sec:denu}, charge and $B'-L'$ neutrality remove $\nu'$ states, leaving $(B'_{ddd}$ + $\bar{e}')$ atoms.  The constraints from self interactions and the ionized fraction of atomic dark matter are evaded if $\alpha_{B-L}' \gtrsim 10^{-2}$.

\subsubsection{Light $u'$, $d'$, $e'$ with no light $B'-L'$ gauge boson: mirror atom and/or neutron}
If $\nu'$ is sufficiently heavy, it decays to $u'$, $d'$ and $e'$ and disappears from the thermal bath.
Reactions mediated by $W'$ are absent, so that $u'$, $d'$ and $e'$ numbers are separately conserved. The following mirror particles are stable,
\begin{align}
B'_{uuu},~B'_{uud},~B'_{udd},~B'_{ddd},~M'_{u\bar{d}},~e'.\nonumber
\end{align}
Among them, $B'_{uuu}$, $B'_{ddd}$ and $M'_{u\bar{d}}$ disappear as discussed in section~\ref{sec:udnu}.
The asymmetry is stored in mirror protons $B'_{uud}$, neutrons $B'_{udd}$, and electrons $e'$, with
abundances that depend on the $B'$ and $L'$ asymmetries.
If $L'=0$, only the mirror neutron remains, whereas
if $B'-L' =0$, as would occur if $B'-L'$ is unbroken during asymmetry genesis, only the mirror proton 
and the mirror electron remain\footnote{%
\small These cases require asymmetry generation after the mirror sphaleron freeze out, as can be achieved in the supersymmetric Twin Higgs model, with Affleck-Dine baryogenesis~\cite{Affleck:1984fy,Dine:1995kz} from the $\bar{u}'\bar{d}' \bar{s}'$ flat direction.}.

If $B'-L'$ symmetry is broken at a sufficiently high scale,  a non-zero $B'-L'$ asymmetry may be generated above the mirror electroweak phase transition temperature.
After the mirror sphaleron process freezes out, the $B'$ and $L'$ asymmetries are given by
\begin{align}
B' = \frac{1}{4}(B-L)',~~L' = - \frac{3}{4}( B-L)'.
\end{align}
Here we assume that the matter content of the mirror sector is identical to the standard model plus three right-handed neutrinos just before the mirror sphaleron process freezes out.
The resultant number densities of mirror protons, neutrons and electrons are given by
\begin{align}
n_{e'} = n_{L'} = - \frac{3}{4} n_{B'-L'} = - \frac{3}{4}n_{n'},~~
n_{p'} = n_{e'} =  -\frac{3}{4}n_{n'},~~
n_{n'} = n_{B'-L'}
\end{align}
Dark matter is composed of comparable numbers of mirror atoms and mirror neutrons.
The constraints from self interactions and ionized fraction of atomic dark matter are relaxed.

\section{Conclusions}

The Twin Higgs mechanism significantly relaxes fine-tuning of the electroweak scale, and allows for a larger cut off scale.
The cut off of the Standard Model is
\begin{align}
\Lambda_{\rm SM} \simeq 1.4~{\rm TeV}\times \left(\frac{\Delta_{\rm SM}}{10}\right)^{1/2},
\end{align}
while that of the Twin Higgs theory is
\begin{align}
\Lambda_{\rm TH} = 5.7~{\rm TeV} \times \left(\frac{\Delta_{\rm TH}}{10}\right)^{1/2} \lambda^{1/2}.
\end{align}
The minimal theory has no new colored states to be produced at the LHC.  It does offer the possibility of discovery modes at the LHC, such as production of the mirror Higgs via Higgs mixing; but the larger cut off
may raise the masses of new particles above the LHC reach.

Mirror Twin Higgs models, however, predict the existence of extremely light particles, mirror photons and mirror neutrinos, that
contribute to the dark radiation of the universe, leading to constraints on a realistic theory.
We have found that, independent of the interactions that couple the two sectors,  it is {\it necessary} to break the mirror $Z_2$ symmetry in the Yukawa couplings.

\noindent \textbf{Minimal Mirror Twin Higgs}   \hspace{0.25in} We have constructed a completely realistic effective field theory of Twin Higgs below the cut off $\Lambda_{\rm TH}$.
It contains a complete mirror sector, so that a UV completion, which we did not study, can restore spacetime parity symmetry.
In Minimal Mirror Twin Higgs, the only $Z_2$ breaking arises from Yukawa couplings and the only communication between the sectors is from Higgs mixing, required by Twin Higgs, and kinetic mixing of hypercharge fields, allowed by gauge invariance.   Furthermore, the $Z_2$ breaking Yukawa couplings not only suppress dark radiation to acceptable levels, but generate the $Z_2$ breaking Higgs mass term necessary for the Twin Higgs mechanism and raise the mirror baryon mass as required for realistic dark matter.

The $Z_2$ breaking Yukawa couplings induce a sizable invisible branching ratio of the SM-like Higgs boson through its mixing with the mirror Higgs.
This, together with the reduction of the Higgs coupling to Standard Model particles, leads to a universal deviation from unity of the Higgs signal-strengths correlated with the masses of mirror fermions, as shown in Fig.~\ref{fig:mass_mu}.
Irrespective of the mirror fermions masses,
the high-luminosity running of the LHC and the ILC can probe $v'/v<4$ and $10$ respectively.

As illustrated in Fig.~\ref{fig:THiggs}, it is non-trivial that Higgs mixing can lead to a decoupling temperature less than the QCD$'$ phase transition temperature, necessary for a solution of the dark radiation problem.  Fig.~\ref{fig:THiggs} shows that, with Higgs mixing alone, there must be light mirror fermions with mass $m_{f'}$ in the range of about $(4-28)$ GeV.   The upper bound on $m_{f'}$ gives a lower bound of $\Delta N_{\rm eff}\gtrsim 0.3$.  The allowed range for $m_{f'}$ depends on $N_{f'}$, the number of light mirror fermion states, and narrows considerably for larger values of $N_{f'}$.  As the upper bound on $m_{f'}$ becomes tighter, $\Delta N_{\rm eff}$ increases, which is shown in Fig.~\ref{fig:Neff Higgs}.  

The lower bounds on $m_{f'}$ from Figs.~\ref{fig:THiggs} and \ref{fig:Neff Higgs} imply lower bounds on the invisible branching ratio of the SM-like Higgs boson and the universal deviation from unity of the Higgs signal-strengths.
For $v'/v=3$, the left panel of Fig.~\ref{fig:Brinv} shows that this branching ratio is typically in the range of 0.05-0.15 that can be probed by high luminosity running of the LHC.  For $v'/v=5$, the invisible branching ratio is reduced; the right panel of Fig.~\ref{fig:Brinv} shows that in some cases the signal is as small as 0.002 - 0.01, that could be probed by ILC. Essentially the entire parameter space can be probed. 

If the kinetic mixing parameter $\epsilon$ is large enough to affect the thermal history of the two sectors near the QCD$'$ phase transition, the allowed range of the light mirror fermion masses is enlarged.  In section \ref{sec:mixing} we showed that the upper bound on $m_{f'}$ could be extended as high as 50 GeV, while mirror leptons could be as light as 2 GeV.  A mirror quark can be lighter than 2 GeV provided the lightest mirror meson is heavier than 2 GeV.  Even including kinetic mixing, the lower bound on $\Delta N_{\rm eff}$ remains about 0.3, but the enlarged range of light mirror fermion masses is important for mirror dark matter.

Mirror baryons and leptons are natural candidates for dark matter.   Dark matter can be composed of mirror neutrons, mirror atoms, or even a mixture of the two.  Such dark matter can be directly detected via Higgs exchange, as illustrated in Fig. \ref{fig:NNp scat}.  For low $v'/v$, PandaX and LUX have recently excluded large values for the dark matter mass.   In the absence of kinetic mixing, the region of dark matter masses allowed by present limits on dark radiation can be fully explored, up to the $1(2) \sigma$ limit by the XENON1T (LZ) experiment.   Adding kinetic mixing to the thermal cosmological history, the limit on $\Delta N_{\rm eff}$ is consistent with lighter dark masses that XENON1T and LZ are unable to probe.  For mirror neutron dark matter masses in the (1-10) GeV region, present direct detection limits bound $\epsilon < 10^{-3} - 10^{-4}$ from scattering via a dipole moment. Mirror atomic dark matter with a sizable kinetic mixing such that the thermal history is affected is excluded.

In the Minimal Mirror Twin Higgs the seesaw mechanism yields light neutrino masses for both sectors.  These neutrinos can be Majorana with those in the mirror sector heavier by a factor $(v'/v)^2$ than the observed neutrinos, leading to important effects in both CMB and LSS.  The effective sum on neutrino masses relevant for cosmological data is at least 2.3 times greater than in the Standard Model.  Small mixing between these Majorana standard and mirror neutrinos could lead to mirror neutrinos being observed as massive sterile neutrinos.  Alternatively the seesaw could lead to Dirac neutrinos, with the mirror states as right-handed neutrinos degenerate with the observed left-handed states, leading to only very small effects on CMB and LSS.  The predictions of Eqs.~(\ref{eq:mnuM}, \ref{eq:mnuD}) for the masses of mirror neutrinos, however, rely on the assumption that $Z_2$ breaking does not substantially affect either the neutrino Yukawa couplings or right-handed neutrino masses, and therefore our predictions for the effective sum of neutrino masses,
$\left(\sum m_\nu \right)_{\rm eff}$,
are less robust than the predictions for $\Delta N_{\rm eff}$, $\Gamma(h \rightarrow \rm inv)$ and the mirror baryon dark matter direct detection cross section.

\noindent \textbf{Additional $Z_2$ breaking from $B-L$ vevs} \hspace{0.25in}
A large $B-L$ vev can implement the seesaw mechanism for the known neutrinos, while a small or zero $B' - L'$ vev can lead to large Dirac mirror neutrino masses.  Without light mirror neutrinos, the minimal $\Delta N_{\rm eff}$ is lowered to 0.08 (0.17) without (with) a massless $B'-L'$ gauge field.  Furthermore, there are new possibilities for mirror dark matter.  If there is a massless $B'-L'$ gauge boson, dark matter could be $B'_{udd}\nu'$ or $B'_{ddd} \bar{e}'$ atoms, and constraints from self interactions and the ionized fraction are weekend if $\alpha'_{B-L} \gsim \alpha$.  Without a massless $B'-L'$ gauge boson, dark matter could be a mixture of 
$B'_{uud}e'$ atoms and mirror neutrons $B'_{udd}$;
constraints from self interactions and the ionized fraction are relaxed.

\section*{Acknowledgments}
The work of L.~J.~Hall and K.~Harigaya  was supported in part by the Department of Energy, Office of Science, Office of High Energy Physics, under contract No.~DE-AC02-05CH11231, and by the National Science Foundation under grants PHY-1316783 and PHY-1521446.   R.~Barbieri wants to thank Dr. Max R\"ossler, the Walter Haefner Foundation and the ETH Zurich Foundation for support.
 The work of L.~J.~Hall was performed in part at the Institute for Theoretical Studies ETH and at the Aspen Center for Physics, which is supported by National Science Foundation grant PHY-1066293.

\appendix

\section{Heavy axion and  $Z_2$ symmetry breaking in Froggatt Nielsen sector}
One of the advantages of the Mirror World is the possibility of a heavy visible QCD axion~\cite{Rubakov:1997vp,Berezhiani:2000gh,Hook:2014cda,Fukuda:2015ana}.
If the Standard Model and mirror sectors have a common Peccei-Quinn symmetry, the axion mass is given by the mirror QCD dynamics and becomes heavier.
The constraint from colliders and astrophsyics can be evaded even if the Peccei-Quinn symmetry breaking scale is around the TeV scale.
If the physical theta angles of mirror QCD and QCD are identical, $\bar{\theta'} = \bar{\theta}$, the theory still solves the strong CP problem.
Due to the small Peccei-Quinn symmetry breaking scale and the large axion mass, the Peccei-Quinn symmetry can be easily understood as an accidental symmetry~\cite{Rubakov:1997vp,Fukuda:2015ana}.
(See~\cite{Barr:1992qq,Holman:1992us,Dine:1992vx,Dias:2002gg,Carpenter:2009zs,Harigaya:2013vja} for models of an invisible QCD axion with an accidental Peccei-Quinn symmetry.)

Is it not clear whether this idea can be incorporated into models with Mirror Twin Higgs.
As we have shown in this paper, the $Z_2$ symmetry breaking in the Yukawa couplings is mandatory.
This may lead to $\bar{\theta'} \neq \bar{\theta}$, ruining the Peccei-Quinn solution to the strong CP problem~\cite{Albaid:2015axa}.
Here we show that if the $Z_2$ symmetry is broken by VEVs of FN symmetry breaking fields we regain $\bar{\theta'} = \bar{\theta}$.
We note that this mechanism is not peculiar to Twin Higgs models, but is generically applicable to Mirror World scenarios.

Let us denote the FN symmetry breaking field as $\phi$. The determinants of the mass matrix of $SU(3)_c$ and $SU(3)_c'$ charged fermions are proportional to
\begin{align}
{\rm det}_{SU(3)_c}M \propto \phi^A f(|\phi|),~~{\rm det}_{SU(3)_c'}M' \propto \phi^{'A} f(|\phi'|),
\end{align}
where $A$ is the anomaly coefficient of (FN symmetry)-$SU(3)_c$-$SU(3)_c$, and $f$ is a function.
Note that the phases of the mass matrices do not depend on the absolute value of the VEVs of $\phi$ and $\phi'$.
The difference of the FN symmetry breaking scales itself does not ruin the axion solution to the strong CP problem.

If the FN symmetry has no color anomaly (i.e.~$A=0$), 
the determinants of the mass matrices have the same phases for any phases of the $\phi$ and $\phi'$ VEVs, giving $\bar{\theta'} = \bar{\theta}$.  If the FN symmetry has a color anomaly, the theta angles may or may not be identical, depending on how the phases of the VEVs of $\phi$ and $\phi'$ are determined.
For example, suppose that the phase directions of $\phi$ and $\phi'$ are determined by explicit breaking of the continuous FN symmetry to a discrete $Z_N$ subgroup, so that vacua are given by
\begin{align}
\vev{\phi} = |\vev{\phi}|\times {\rm exp}(2\pi i \frac{k}{N} ),~~\vev{\phi'} = |\vev{\phi'}| \times {\rm exp}(2\pi i \frac{k'}{N})~~(k,k' = 0,1,\cdots, N-1).
\end{align}
If $A/N$ is an integer, the theta angles remain identical in any vacua. If not, the theta angles remain identical in specific vacua.

Assuming that mirror quark masses are larger than the dynamical scale of mirror QCD,
the mass of the QCD axion is given by
\begin{align}
m_a \simeq m_{\eta'} \left( \frac{\Lambda'_{\rm QCD}}{\Lambda_{\rm QCD}} \right)^2 \frac{f_\pi}{f_a} \simeq 1~{\rm MeV}\times \left( \frac{\Lambda'_{\rm QCD}/\Lambda_{\rm QCD}}{10} \right)^2 \frac{10~{\rm TeV}}{f_a}.
\end{align}
Comparing this formula with Figs.~1 and 2 of~\cite{Fukuda:2015ana},  constraints from colliders and $\Delta N_{\rm eff}$ are satisfied for $f_a\gtrsim 10$ TeV or $f_a \lesssim 100$ GeV.
For $f_a\gtrsim 10$ TeV, however, the QCD axion decays into photons and mirror photons after BBN begins, giving a slight difference between the baryon-to-photon ratio during BBN and during recombination.
To conclude whether such a possibility is excluded or not, a calculation of BBN with decaying QCD axions,
as well as a calculation of the CMB spectrum including the effect of $\Delta N_{\rm eff}$ and non-zero neutrino masses, are required,
which is beyond the scope of this paper.
For $f_a \lesssim 100$ GeV, there should be $SU(3)_c$ particles with masses of $O(100)$ GeV, which is excluded by hadron colliders%
\footnote{\small If the Peccei-Quinn symmetry is broken by strong dynamics, $SU(3)_c$ charged fermions may be as heavy as $O(1)$ TeV. However, it is difficult to avoid light $SU(3)_c$ charged pseudo Nambu-Goldstone bosons. If the domain wall number is $O(10)$, all $SU(3)_c$ charged particles can be as heavy as $O(1)$ TeV.}.

\begingroup
\renewcommand{\addcontentsline}[3]{}
\renewcommand{\section}[2]{}

\endgroup
  

\begin{thebibliography}{99} 

\bibitem{Lee:1956qn} 
  T.~D.~Lee and C.~N.~Yang,
  Phys.\ Rev.\  {\bf 104}, 254 (1956).

\bibitem{Kobzarev:1966qya} 
  I.~Y.~Kobzarev, L.~B.~Okun and I.~Y.~Pomeranchuk,
  Sov.\ J.\ Nucl.\ Phys.\  {\bf 3}, no. 6, 837 (1966)
  [Yad.\ Fiz.\  {\bf 3}, 1154 (1966)].

\bibitem{Chacko:2005pe} 
  Z.~Chacko, H.~S.~Goh and R.~Harnik,
  Phys.\ Rev.\ Lett.\  {\bf 96}, 231802 (2006)
  [hep-ph/0506256].
  
\bibitem{Barbieri:2005ri} 
  R.~Barbieri, T.~Gregoire and L.~J.~Hall,
  hep-ph/0509242.

\bibitem{Chang:2006ra} 
  S.~Chang, L.~J.~Hall and N.~Weiner,
  Phys.\ Rev.\ D {\bf 75}, 035009 (2007)
  [hep-ph/0604076].

\bibitem{Craig:2013fga} 
  N.~Craig and K.~Howe,
  JHEP {\bf 1403}, 140 (2014)
  [arXiv:1312.1341 [hep-ph]].

\bibitem{Batra:2008jy} 
  P.~Batra and Z.~Chacko,
  Phys.\ Rev.\ D {\bf 79}, 095012 (2009)
  [arXiv:0811.0394 [hep-ph]].


\bibitem{Geller:2014kta} 
  M.~Geller and O.~Telem,
  Phys.\ Rev.\ Lett.\  {\bf 114}, 191801 (2015)
  [arXiv:1411.2974 [hep-ph]].

\bibitem{Barbieri:2015lqa} 
  R.~Barbieri, D.~Greco, R.~Rattazzi and A.~Wulzer,
  JHEP {\bf 1508}, 161 (2015)
  [arXiv:1501.07803 [hep-ph]].

\bibitem{Low:2015nqa} 
  M.~Low, A.~Tesi and L.~T.~Wang,
  Phys.\ Rev.\ D {\bf 91}, 095012 (2015)
  [arXiv:1501.07890 [hep-ph]].
  
\bibitem{Buttazzo:2015bka} 
  D.~Buttazzo, F.~Sala and A.~Tesi,
  JHEP {\bf 1511}, 158 (2015)
  [arXiv:1505.05488 [hep-ph]].


\bibitem{Craig:2015pha} 
  N.~Craig, A.~Katz, M.~Strassler and R.~Sundrum,
  JHEP {\bf 1507}, 105 (2015)
  [arXiv:1501.05310 [hep-ph]].

\bibitem{Khachatryan:2016vau} 
  G.~Aad {\it et al.} [ATLAS and CMS Collaborations],
  arXiv:1606.02266 [hep-ex].


\bibitem{Manohar:1983md} 
  A.~Manohar and H.~Georgi,
  Nucl.\ Phys.\ B {\bf 234}, 189 (1984).
  
\bibitem{Luty:1997fk} 
  M.~A.~Luty,
  Phys.\ Rev.\ D {\bf 57}, 1531 (1998)
  [hep-ph/9706235].

\bibitem{'tHooft:1973jz} 
  G.~'t Hooft,
  Nucl.\ Phys.\ B {\bf 72}, 461 (1974).

\bibitem{Borsanyi:2016ksw} 
  S.~Borsanyi {\it et al.},
  arXiv:1606.07494 [hep-lat].

\bibitem{Yanagida:1979as} 
  T.~Yanagida,
  Conf.\ Proc.\ C {\bf 7902131}, 95 (1979).

\bibitem{GellMann:1980vs} 
  M.~Gell-Mann, P.~Ramond and R.~Slansky,
  Conf.\ Proc.\ C {\bf 790927}, 315 (1979)
  [arXiv:1306.4669 [hep-th]].

\bibitem{Minkowski:1977sc} 
  P.~Minkowski,
  Phys.\ Lett.\ B {\bf 67}, 421 (1977).

\bibitem{Ade:2015xua} 
  P.~A.~R.~Ade {\it et al.} [Planck Collaboration],
  arXiv:1502.01589 [astro-ph.CO].

\bibitem{Froggatt:1978nt} 
  C.~D.~Froggatt and H.~B.~Nielsen,
  Nucl.\ Phys.\ B {\bf 147}, 277 (1979).

\bibitem{ATLAS_prospect}
ATLAS Collaboration,
ATLAS-PHYS-PUB-2014-017.

\bibitem{Asner:2013psa} 
  D.~M.~Asner {\it et al.},
  arXiv:1310.0763 [hep-ph].


\bibitem{Okamoto:1999hi} 
  M.~Okamoto {\it et al.} [CP-PACS Collaboration],
  Phys.\ Rev.\ D {\bf 60}, 094510 (1999)
  [hep-lat/9905005].

\bibitem{Borsanyi:2012ve} 
  S.~Borsanyi, G.~Endrodi, Z.~Fodor, S.~D.~Katz and K.~K.~Szabo,
  JHEP {\bf 1207}, 056 (2012)
  [arXiv:1204.6184 [hep-lat]].


\bibitem{CMS:2013xfa} 
  [CMS Collaboration],
  arXiv:1307.7135.


\bibitem{Davidson:1991si} 
  S.~Davidson, B.~Campbell and D.~C.~Bailey,
  Phys.\ Rev.\ D {\bf 43}, 2314 (1991).

\bibitem{Jaeckel:2012yz} 
  J.~Jaeckel, M.~Jankowiak and M.~Spannowsky,
  Phys.\ Dark Univ.\  {\bf 2}, 111 (2013)
  [arXiv:1212.3620 [hep-ph]].


\bibitem{Haas:2014dda} 
  A.~Haas, C.~S.~Hill, E.~Izaguirre and I.~Yavin,
  Phys.\ Lett.\ B {\bf 746}, 117 (2015)
  [arXiv:1410.6816 [hep-ph]].


\bibitem{Goldberg:1986nk} 
  H.~Goldberg and L.~J.~Hall,
  Phys.\ Lett.\ B {\bf 174}, 151 (1986).

\bibitem{Blinnikov:1982eh} 
  S.~I.~Blinnikov and M.~Y.~Khlopov,
  Sov.\ J.\ Nucl.\ Phys.\  {\bf 36}, 472 (1982)
  [Yad.\ Fiz.\  {\bf 36}, 809 (1982)];
  Sov.\ Astron.\  {\bf 27}, 371 (1983)
  [Astron.\ Zh.\  {\bf 60}, 632 (1983)].

\bibitem{Kolb:1985bf} 
  E.~W.~Kolb, D.~Seckel and M.~S.~Turner,
  Nature {\bf 314}, 415 (1985).


\bibitem{Sommerfeld}
A.~Sommerfeld, Annalen der Physik {\bf 403}, 257 (1931).

\bibitem{Hisano:2002fk}
J.~Hisano, S.~Matsumoto and M.~M.~Nojiri,
Phys.\ Rev.\ D {\bf 67}, 075014 (2003)
[hep-ph/0212022].

\bibitem{Cline:2013zca} 
  J.~M.~Cline, Z.~Liu, G.~Moore and W.~Xue,
  Phys.\ Rev.\ D {\bf 90}, no. 1, 015023 (2014)
  doi:10.1103/PhysRevD.90.015023
  [arXiv:1312.3325 [hep-ph]].


\bibitem{Cline:2013pca} 
  J.~M.~Cline, Z.~Liu, G.~Moore and W.~Xue,
  Phys.\ Rev.\ D {\bf 89}, no. 4, 043514 (2014)
  [arXiv:1311.6468 [hep-ph]].

\bibitem{Kaplinghat:2015aga} 
  M.~Kaplinghat, S.~Tulin and H.~B.~Yu,
  Phys.\ Rev.\ Lett.\  {\bf 116}, no. 4, 041302 (2016)
  [arXiv:1508.03339 [astro-ph.CO]].
  

\bibitem{CyrRacine:2012fz} 
  F.~Y.~Cyr-Racine and K.~Sigurdson,
  Phys.\ Rev.\ D {\bf 87}, no. 10, 103515 (2013)
  [arXiv:1209.5752 [astro-ph.CO]].

\bibitem{Dubovsky:2001tr} 
  S.~L.~Dubovsky and D.~S.~Gorbunov,
  Phys.\ Rev.\ D {\bf 64}, 123503 (2001)
  [astro-ph/0103122].

\bibitem{Dubovsky:2003yn} 
  S.~L.~Dubovsky, D.~S.~Gorbunov and G.~I.~Rubtsov,
  JETP Lett.\  {\bf 79}, 1 (2004)
  [Pisma Zh.\ Eksp.\ Teor.\ Fiz.\  {\bf 79}, 3 (2004)]
  [hep-ph/0311189].

\bibitem{Dolgov:2013una} 
  A.~D.~Dolgov, S.~L.~Dubovsky, G.~I.~Rubtsov and I.~I.~Tkachev,
  Phys.\ Rev.\ D {\bf 88}, no. 11, 117701 (2013)
  [arXiv:1310.2376 [hep-ph]].

\bibitem{Chuzhoy:2008zy} 
  L.~Chuzhoy and E.~W.~Kolb,
  JCAP {\bf 0907}, 014 (2009)
  [arXiv:0809.0436 [astro-ph]].

\bibitem{Craig:2015xla} 
  N.~Craig and A.~Katz,
  JCAP {\bf 1510}, no. 10, 054 (2015)
  [arXiv:1505.07113 [hep-ph]].

\bibitem{Garcia:2015toa} 
  I.~Garcia Garcia, R.~Lasenby and J.~March-Russell,
  Phys.\ Rev.\ Lett.\  {\bf 115}, no. 12, 121801 (2015)
  [arXiv:1505.07410 [hep-ph]].


\bibitem{Farina:2015uea} 
  M.~Farina,
  JCAP {\bf 1511}, no. 11, 017 (2015)
  [arXiv:1506.03520 [hep-ph]].



\bibitem{Djouadi:2000ck} 
  A.~Djouadi and M.~Drees,
  Phys.\ Lett.\ B {\bf 484}, 183 (2000)
  [hep-ph/0004205].

\bibitem{Shifman:1978zn} 
  M.~A.~Shifman, A.~I.~Vainshtein and V.~I.~Zakharov,
  Phys.\ Lett.\ B {\bf 78}, 443 (1978).

\bibitem{Abdel-Rehim:2016won} 
  A.~Abdel-Rehim {\it et al.} [ETM Collaboration],
  Phys.\ Rev.\ Lett.\  {\bf 116}, no. 25, 252001 (2016)
  [arXiv:1601.01624 [hep-lat]].

\bibitem{Alarcon:2011zs} 
  J.~M.~Alarcon, J.~Martin Camalich and J.~A.~Oller,
  Phys.\ Rev.\ D {\bf 85}, 051503 (2012)
  [arXiv:1110.3797 [hep-ph]].

\bibitem{Alarcon:2012nr} 
  J.~M.~Alarcon, L.~S.~Geng, J.~Martin Camalich and J.~A.~Oller,
  Phys.\ Lett.\ B {\bf 730}, 342 (2014)
  [arXiv:1209.2870 [hep-ph]].

\bibitem{Akerib:2016vxi} 
  D.~S.~Akerib {\it et al.},
  arXiv:1608.07648 [astro-ph.CO].

\bibitem{Tan:2016zwf} 
  A.~Tan {\it et al.} [PandaX-II Collaboration],
  arXiv:1607.07400 [hep-ex].

\bibitem{Aprile:2015uzo} 
  E.~Aprile {\it et al.} [XENON Collaboration],
  JCAP {\bf 1604}, no. 04, 027 (2016)
  [arXiv:1512.07501 [physics.ins-det]].

\bibitem{Akerib:2015cja} 
  D.~S.~Akerib {\it et al.} [LZ Collaboration],
  arXiv:1509.02910 [physics.ins-det].

\bibitem{Billard:2013qya} 
  J.~Billard, L.~Strigari and E.~Figueroa-Feliciano,
  Phys.\ Rev.\ D {\bf 89}, no. 2, 023524 (2014)
  [arXiv:1307.5458 [hep-ph]].

\bibitem{DelNobile:2014eta} 
  E.~Del Nobile, G.~B.~Gelmini, P.~Gondolo and J.~H.~Huh,
  JCAP {\bf 1406}, 002 (2014)
  [arXiv:1401.4508 [hep-ph]].

\bibitem{Angloher:2015eza} 
G.~Angloher {\it et al.} [CRESST Collaboration],
  arXiv:1503.08065 [astro-ph.IM].

\bibitem{Calkins:2016pnm} 
  R.~Calkins [SuperCDMS Collaboration],
  J.\ Phys.\ Conf.\ Ser.\  {\bf 718}, no. 4, 042009 (2016).

\bibitem{Cline:2012is} 
  J.~M.~Cline, Z.~Liu and W.~Xue,
  Phys.\ Rev.\ D {\bf 85}, 101302 (2012)
  [arXiv:1201.4858 [hep-ph]].

\bibitem{Angloher:2015ewa} 
  G.~Angloher {\it et al.} [CRESST Collaboration],
  Eur.\ Phys.\ J.\ C {\bf 76}, no. 1, 25 (2016)
  [arXiv:1509.01515 [astro-ph.CO]].

\bibitem{Affleck:1984fy} 
  I.~Affleck and M.~Dine,
  Nucl.\ Phys.\ B {\bf 249}, 361 (1985).

\bibitem{Dine:1995kz} 
  M.~Dine, L.~Randall and S.~D.~Thomas,
  Nucl.\ Phys.\ B {\bf 458}, 291 (1996)
  [hep-ph/9507453].

\bibitem{Rubakov:1997vp} 
  V.~A.~Rubakov,
  JETP Lett.\  {\bf 65}, 621 (1997)
  [hep-ph/9703409].

\bibitem{Berezhiani:2000gh} 
  Z.~Berezhiani, L.~Gianfagna and M.~Giannotti,
  Phys.\ Lett.\ B {\bf 500}, 286 (2001)
  [hep-ph/0009290].


\bibitem{Hook:2014cda} 
  A.~Hook,
  Phys.\ Rev.\ Lett.\  {\bf 114}, no. 14, 141801 (2015)
  [arXiv:1411.3325 [hep-ph]].
  
\bibitem{Fukuda:2015ana} 
  H.~Fukuda, K.~Harigaya, M.~Ibe and T.~T.~Yanagida,
  Phys.\ Rev.\ D {\bf 92}, no. 1, 015021 (2015)
  [arXiv:1504.06084 [hep-ph]].


\bibitem{Barr:1992qq} 
  S.~M.~Barr and D.~Seckel,
  Phys.\ Rev.\ D {\bf 46}, 539 (1992).

\bibitem{Holman:1992us} 
  R.~Holman, S.~D.~H.~Hsu, T.~W.~Kephart, E.~W.~Kolb, R.~Watkins and L.~M.~Widrow,
  Phys.\ Lett.\ B {\bf 282}, 132 (1992)
  [hep-ph/9203206].

\bibitem{Dine:1992vx} 
  M.~Dine,
  hep-th/9207045.

\bibitem{Dias:2002gg} 
  A.~G.~Dias, V.~Pleitez and M.~D.~Tonasse,
  Phys.\ Rev.\ D {\bf 67}, 095008 (2003)
  [hep-ph/0211107].

\bibitem{Carpenter:2009zs} 
  L.~M.~Carpenter, M.~Dine and G.~Festuccia,
  Phys.\ Rev.\ D {\bf 80}, 125017 (2009)
  [arXiv:0906.1273 [hep-th]].

\bibitem{Harigaya:2013vja} 
  K.~Harigaya, M.~Ibe, K.~Schmitz and T.~T.~Yanagida,
  Phys.\ Rev.\ D {\bf 88}, no. 7, 075022 (2013)
  [arXiv:1308.1227 [hep-ph]];
  Phys.\ Rev.\ D {\bf 92}, no. 7, 075003 (2015)
  [arXiv:1505.07388 [hep-ph]].

\bibitem{Albaid:2015axa} 
  A.~Albaid, M.~Dine and P.~Draper,
  JHEP {\bf 1512}, 046 (2015)
  [arXiv:1510.03392 [hep-ph]].

\end{thebibliography}
\end{document}